\documentstyle[12pt, slashbox]{article}
\def\BorL{b }	
\makeatletter
\newif\ifpr@pstyle \pr@pstylefalse
\newif\ifnons@qeq  \nons@qeqfalse
\def\bigpage{
	\setlength{\topmargin}{-.5in}
	\setlength{\oddsidemargin}{.5pc}
	\setlength{\evensidemargin}{.5pc}
	\setlength{\textwidth}{35pc}
	\setlength{\textheight}{56pc}
	\setlength{\parskip}{6pt plus 2pt minus 1pt}
	\newlength{\paperbaselineskip}
	\setlength{\paperbaselineskip}{20pt plus 0.2pt minus 0.1pt}
	\def\@oddfoot{\hfil -- \thepage~--\hfil}
	\let\@evenfoot\@oddfoot
        \def\thesection{\arabic{section}.}
        \def\thesubsection{\thesection\arabic{subsection}}
        \def\@ourappendix{\par\setcounter{section}{0}
                      \setcounter{subsection}{0}
                      \def\thesection{\Alph{section}.}
                      \ifnons@qeq
                      \def\theequation{\Alph{section}.\arabic{equation}}\fi}
        \def\appendix{\@ourappendix}
        \def\section{\@startsection {section}{1}%
            {\z@}{5ex plus .2ex minus .4ex}%
            {1.5ex plus.4ex minus .1ex}%
            {\centering\ifpr@pstyle\else\ifx\undefined\reset@font\else%
             \reset@font\fi\large\fi\bf}}
        \def\subsection{\@startsection{subsection}%
            {2}{\z@}{3.25ex plus .4ex minus .4ex}%
            {1ex plus .2ex}{\bf}}
}
\bigpage
\newfont{\fourteencp}{cmcsc10 scaled\magstep2}
\newfont{\titlefont}{cmbx10 scaled\magstep2}
\newfont{\authorfont}{cmcsc10 scaled\magstep1}
\newfont{\fourteenmib}{cmmib10 scaled\magstep2}
	\skewchar\fourteenmib='177
\newfont{\elevenmib}{cmmib10 scaled\magstephalf}
	\skewchar\elevenmib='177
\newfont{\ninemib}{cmmib9} \skewchar\ninemib='177
\makeatletter
\newcommand\nonsequentialeqnum{
        \nons@qeqtrue
	\@addtoreset{equation}{section}
	\def\theequation{\arabic{section}.\arabic{equation}}}
\newif\ifp@bblock  \p@bblocktrue
\newcommand\nopubblock{\p@bblockfalse}
\newcommand\topspace{\hrule height 0pt depth 0pt \vskip}
\newcommand\p@bblock{\begingroup \tabskip=\hsize minus \hsize
	\baselineskip=1.5\ht\strutbox \topspace-2\baselineskip
	\halign to\hsize{\strut ##\hfil\tabskip=0pt\crcr
	\the\Pubnum\crcr\the\date\crcr}\endgroup}
\newcommand\YUKAWAmark{\hbox{
        \ifpr@pstyle\ninemib\else\elevenmib\fi
        Yukawa\hskip1mm Institute\hskip1mm Kyoto \hfill}}
\newtoks\date
\newtoks\Pubnum
\let\pubnum=\Pubnum
\Pubnum={}
\date={\today}
\newcommand{\frontpageskip}{\vspace{12pt plus .5fil minus 2pt}}
\def\@authoraddress{} \def\@title{}
\def\title#1{\gdef\@title{\frontpageskip
	\begin{center}{\titlefont #1}\end{center}\par}}
\def\@author#1{\frontpageskip\par\begin{center}{\authorfont #1}
	\end{center}
	\nobreak}
\def\author#1{\expandafter\def\expandafter\@authoraddress\expandafter
    {\@authoraddress{\@author{#1}}}}
\def\andauthor#1{\expandafter\def\expandafter\@authoraddress\expandafter
    {\@authoraddress{\frontpageskip\centerline{and}\@author{#1}}}}
\def\authors#1{\expandafter\def\expandafter\@authoraddress\expandafter
    {\@authoraddress{\frontpageskip\noindent #1}}}
\def\@address#1{\par\begin{center}{\sl #1}\end{center}\par}
\def\address#1{\expandafter\def\expandafter\@authoraddress\expandafter
    {\@authoraddress{\@address{#1}}}}
\def\andaddress#1{\expandafter\def\expandafter%
    \@authoraddress\expandafter
    {\@authoraddress{\par\centerline{\sl and}\@address{#1}}}}
\renewcommand{\thanks}[1]{\footnote{#1}}
\def\maketitle{\par
  \begingroup
       \def\thefootnote{\fnsymbol{footnote}}
	\thispagestyle{empty}
        \baselineskip=\paperbaselineskip
	\@maketitle
	\endgroup
	\setcounter{footnote}{0}
	\let\maketitle\relax \let\@maketitle\relax
	\let\@thanks\relax \let\@title\relax
	\let\@title\relax \let\@authoraddress\relax
	\let\thanks\relax}
\def\@maketitle{%
        \ifpr@pstyle\vspace{-1.0cm}\else\vspace{-1.7cm}\fi
	\YUKAWAmark\vskip0.6cm
	\ifp@bblock\p@bblock \else\hrule height 0pt \relax \fi
	\@title
	\@authoraddress
	}
\renewcommand{\abstract}{\par\frontpageskip\centerline{
             \ifpr@pstyle\twelvecp\else\fourteencp\fi Abstract}
	\vspace{8pt plus 3pt minus 3pt}}
\@addtoreset{equation}{section}
\def\theequation{\arabic{section}.\arabic{equation}}
\makeatother
\def\bigmode{b }
\ifx\BorL\undefined\read-1 to\BorL\fi
\makeatletter
\def\doublepage{
        \twocolumn
        
        \pr@pstyletrue
        \sloppy
        \flushbottom
        \setlength{\topmargin}{-0.95in}
        \setlength{\headsep}{20pt}
        \setlength{\headheight}{10pt}
        \hoffset=-0.35in
        \leftmargini 2em
        \leftmarginv .5em
        \leftmarginvi .5em
        \marginparwidth 48pt
        \marginparsep 10pt
        \setlength{\columnsep}{0.7truein}
        \setlength{\textwidth}{10.5truein}
        \setlength{\textheight}{7truein}
        \setlength{\oddsidemargin}{0.0truein}
        \setlength{\evensidemargin}{0.0truein}
        \multiply\paperbaselineskip by 4
                   \divide\paperbaselineskip by 5
        \multiply\footskip by 4 \divide\footskip by 5
        \setlength{\parskip}{4pt plus 1.5pt minus 1pt}
        \newlength{\halfwidth}
        \halfwidth=\textwidth\advance\halfwidth by -\columnsep
                         \divide\halfwidth by 2
        \newfont{\twelvemib}{cmmib10 scaled\magstep1}
                 \skewchar\twelvemib='177
        \newfont{\tenmib}{cmmib10}
                 \skewchar\tenmib='177
        \newfont{\twelvecp}{cmcsc10 scaled\magstep1}
        \def\pagebox{\hbox to \halfwidth{\hfil  -- \thepage~--\hfil}}
        \def\@oddfoot{\pagebox\hfil\addtocounter{page}{1}\pagebox}
        \let\@evenfoot\@oddfoot
        \def\ps@empty{\let\@mkboth\@gobbletwo\let\@oddhead\@empty
               \def\@oddfoot{\hbox to \halfwidth{\hfil ~~~~~~~}\hfil
               \addtocounter{page}{1}\pagebox}
                \let\@evenhead\@empty\let\@evenfoot\@oddfoot}
        \def\appendix{\@ourappendix}
        \def\section{\@startsection {section}{1}%
            {\z@}{5ex plus .2ex minus .4ex}%
            {1.5ex plus.4ex minus .1ex}%
            {\centering\ifpr@pstyle\else\reset@font\large\fi\bf}}
        \def\subsection{\@startsection{subsection}%
            {2}{\z@}{3.25ex plus .4ex minus .4ex}%
            {1ex plus .2ex}{\bf}}
}
\ifx\BorL\bigmode
\else
        \input art10.sty
        \doublepage
\fi
\makeatother
\makeatletter
\newif\ifepsfloaded
\newif\iffigureexists

\ifeof 1 \epsfloadedfalse \else \epsfloadedtrue \fi
\ifepsfloaded \input epsf \fi

\def\checkex#1 {\relax
    \openin 1 #1
    \ifeof 1 \figureexistsfalse
    \else \figureexiststrue
    \fi \closein 1 }
\def\figinsertraw#1#2{
   \ifepsfloaded
       \checkex #1
       \iffigureexists
           \immediate\write16{(#1)}
           #2
       \else
           \immediate\write16{(#1 NOT FOUND!)}
           \vbox to 2in{\hbox to 2in {\hss} \vss}
       \fi
   \else
       \immediate\write16{(NOT inputting #1; no epsf.tex)}
       \vbox to 2in{\hbox to 2in {\hss} \vss}
   \fi}
\newcommand{\reduceland}[2]{\dimen@=#1
     \ifpr@pstyle\multiply\dimen@ by 4\divide\dimen@ by 5\fi
     \edef#2{\dimen@}}
\def\F@gin#1#2#3#4{
  \ifepsfloaded
    \checkex #1
    \iffigureexists
        \immediate\write16{(#1)}
        \begin{figure}
        \ifdim#2>\z@\reduceland{#2}{\dimen@ii}\epsfxsize=\dimen@ii\fi
        \ifdim#3>\z@\reduceland{#3}{\dimen@ii}\epsfysize=\dimen@ii\fi
        \centerline{\epsfbox{#1}}
        {#4} \end{figure}
    \else
        \immediate\write16{(#1 NOT FOUND!)}
        \begin{figure}
        \ifdim#2>\z@\reduceland{#2}{\dimen@ii}\else\dimen@ii=2in\fi
        \ifdim#3>\z@\reduceland{#3}{\dimen255}\else\dimen255=2in\fi
        \centerline{\framebox[\dimen@ii]{\rule{0pt}{\dimen255}#1}}
        {#4} \end{figure}
    \fi
  \else
    \immediate\write16{(NOT inputting #1; no epsf.tex)}
    \begin{figure}
    \centerline{\framebox[2in]{\rule{0pt}{2in}#1}}
    #4\end{figure}
  \fi}
\def\figinsertx#1#2#3{\F@gin{#1}{#2}{0pt}{#3}}
\def\figinserty#1#2#3{\F@gin{#1}{0pt}{#2}{#3}}
\def\figinsert#1#2{\F@gin{#1}{0pt}{0pt}{#2}}
\makeatother
\begin{document}    
%
\pubnum{YITP-98-21\cr NBI-HE-98-08}
\date{July 1998}

\title{Dirac Sea for Bosons 
}

\author{Holger B. Nielsen
\thanks{e-mail address hbech@nbivms.nbi.dk}}
\address{Niels Bohr Institute, \\
	University of Copenhagen, 17 Blegdamsvej, \\
	Copenhagen $\o$, DK 2100, Denmark\\
	$\&$ \\
	Yukawa Institute for Theoretical Physics \\
	Kyoto University, ~Sakyo-ku, ~Kyoto 606-8502, ~Japan\\}

\andauthor{Masao Ninomiya
\thanks{e-mail address ninomiya@yukawa.kyoto-u.ac.jp}}
\address{Yukawa Institute for Theoretical Physics\\
        Kyoto University,~Sakyo-ku,~Kyoto 606-8502,~Japan\\}

\maketitle


\begin{abstract}
It is proposed to make formulation of second quantizing a bosonic
theory by generalizing the method of filling the Dirac negative energy 
sea for fermions.
We interpret that the correct vacuum for the bosonic theory is
obtained  by adding minus one boson to each single particle negative
energy states while the positive energy states are empty.
The boson states are divided into two sectors ; 
the usual positive sector with positive and zero numbers of bosons and 
the negative sector with negative numbers of bosons.
Once it comes into the negative sector it cannot return to the usual
positive sector by ordinary interaction due to a barrier.

It is suggested to use as a playround models in which the filling of
empty fermion Dirac sea and the removal of boson from the
negative energy states are not yet performed.
We put forward such a naive vacuum world and propose a CPT-like
theorem for it.
We study it in detail and give a proof for 
$\frac{\lambda}{4}(\varphi^+\varphi)^2$
theory.
The CPT-like theorem is a strong reflection, but does not include
inversion of operator order which is involved in the ordinary CPT
theorem.
Instead it needs certain analytic continuation of the multiple wave 
function when the state is formulated as a finite number of particles
present.

\end{abstract}


\section{Introduction} 

There has been an wellknown method, though not popular nowadays, to second 
quantize relativistic fermion by imagining that there is a priori so
called naive vacuum in which there is no, neither positive energy nor
negative energy, electron present.
However this vacuum is unstable and the negative energy state get
filled whereby the Dirac sea is formed
\cite{pamdirac}\footnote{See for example \cite{weinberg} for
historical account.}.
This method of filling at first empty Dirac sea seems to make sense
only for fermions for which there is Pauli principle.
In this way ``correct vacuum'' is formed out of ``naive vacuum'', the
former well functioning phenomenologically.
Formally by filling Dirac sea we define creation operators
$b^+(\stackrel{\rightharpoonup}{p},s,\omega)$
for holes which is equivalent to destruction operators
$a(-\stackrel{\rightharpoonup}{p},-s,-\omega)$
for negative energies $-\omega$ and altogether opposite quantum
numbers.
This formal rewriting can be used also for bosons, but we have never
heard the associated filling of the negative energy states.

It is the first purpose of the present article to present the method
analogous to the filling of the empty Dirac Sea, naive vacuum, but now
for \underline{bosons}.
Unlike to the fermion case, for bosons we must add minus one particle
to each negative energy single particle states.
This may sound curious and is of course formal.
However in the following section 2 we shall treat the harmonic
oscillator which is brought in correspondence with a single particle
state under the usual second quantization.
We make an extension of the spectrum with the excitation number
$n=0,1,2,\ldots$ to the one with negative integer values
$n=-1,-2.\ldots$.
This extension can be performed by requiring that the wave function
$\psi(x)$ should be analytic in the whole complex $x$ plane except for 
an essential singularity at $x=\infty$. 
This requirement is a replacement to the usual condition on the norm
of the finite Hilbert space
$\int^\infty_{-\infty}|\psi(x)|^2d{\rm x}<\infty$.
The outcome of the study is that the harmonic oscillator has the
following two sectors : 
1) the usual positive sector with positive and zero numbers of
particles, and 
2) the negative sector with the negative number of particles.
The 2nd sector has indefinite Hilbert product.

But we would like to stress that there is a barrier between the usual
positive sector and the negative sector.
Due to the barrier it is impossible to pass from one sector to the other
with usual polynomial interactions.
This is due to some extrapolation of the
wellknown laser effect, which make easy to fill an already highly
filled single particle state for bosons.
This laser effect may become zero when an interaction tries to have
the number of particles pass the barrier. 
In this way we may explain that the barrier prevents us from observing
a negative number of bosons.

Once the barrier is passed by removing one particle, the barrier
prevents that there ever again be a non-negative number of particles
in that state.
Contrary to the fermion case it is impossible to remove one particle
from each negative energy boson state.

It may be possible to use as a playground a formal world in which
one has neither yet filled the usual Dirac sea of fermions nor
performed the one boson removal from the negative energy state.
We shall indeed study such a playground model referred to as the naive 
vacuum model.
Particularly we shall provide an analogous theorem to the CPT
theorem 
\footnote{The CPT theorem is well explained in \cite{sakurai}},
since the naive vacuum is \underline{not} CPT invariant for
both fermions and bosons.
At first one might think that a strong reflection without associated
inversion of operator order might be good enough.
But it turns out this has the unwanted feature that the sign of
the interaction energy is not changed. 
This changing the sign is required since under strong reflection the
sign of all energies should be switched.
To overcome this problem we propose the CPT-like symmetry for the
naive vacuum world to include further a certain analytic continuation.
This is constructed by applying a certain analytic continuation around 
a branch point which appears in the wave function for each
pair of particles.
It is presupposed that we can restrict our attention to such a family
of wave function as the one with sufficiently good physical properties.

The second purpose of the present paper, which is the major one, 
is to put forward a physical picture that may be of value in 
developing an intuition on naive vacuum world.
In fact investigation of naive vacuum world may be very attractive
because the physics there is quantum mechanics of finite number of
particles. 
Furthermore the theory is piecewise free in the sense that due to
relativistic interactions become of infinitely short range.
Thus the support that there are interactions is null set and one may
say that the theory is free almost everywhere.
But the very local interactions make themselves felt only via boundary
conditions where two or more particles meet.
This makes the naive vacuum world a theoretical playground. 
However it suffers from the following severe drawbacks from a
physical point of view :

\begin{itemize}
\item No bottom in the Hamiltonian
\item Negative norm square states
\item Pairs of particles with tachyonically moving center of mass
\item It is natural to work with ``anti-bound states'' rather that
bound states in the negative energy regime.
\end{itemize}

What we really want to present is a more dramatic formulation of
relativistic second quantization and one may think of it as a
quantization procedure.
We shall formulate below the shift of vacuum for bosons as a shift of
boundary conditions in the wave functional formulation of the second
quantized theory.

But, using the understanding of second quantization of particles along 
the way we describe, could we get a better understanding of how to
second quantize strings?
This is our original motivation of the present work.
In the oldest attempt to make string field theory by Kaku and
Kikkawa \cite{kaku} the infinite momentum frame was used.
To us it looks like an attempt to escape the problem of the negative
energy states.
But this is the root of the trouble to be resolved by the modification 
of the vacuum ; filling with fermion and removal of boson from each 
negative energy state.
So the hope would be that by grasping better these Dirac sea problems
in our way, one might get the possibility of inventing new types of
string field theories, where the infinite momentum frame would not be
called for. 
We have in mind a string field theory of the type that infinitesimal
pieces rather than full strings
\footnote{Some works along this line are listed in 
\cite{witten},\cite{bergman}}
are second quantized : in such
attempts the Dirac sea of negative energy problems would occur locally 
all along the string.
There seems to be a possibility to escape in some way the infinite momentum
frame in string filed theory.
The real problem is that some splitting of the string so as to avoid
the duality equality of $s-$ and $t-$ channels may be needed instead
of the summation of the Feynmann diagrams.
To avoid such problems a quantization of constituents may look more
promising.
But then the negative energy may occur for pieces of strings and the
problem may be further enhanced.
Thus at first we might be happy to make a string field theory in the
formulation of the naive vacuum world.

The present paper is organized as follows. 
Before going to the real description of how to quantize bosons in our
formulation we shall formally look at the
harmonic oscillator in the following section 2.
It is naturally extended to describe a single particle state that can
also have a negative number of particles in it. 
In section 3 application to even spin particles is described, where
the negative norm square problems are gotten rid of.
In section 4 we bring our method into a wave functional formulation,
wherein changing the convergence and finite norm 
condition are explained.
In section 5 we illustrate the main point of the formulation of the
wave functional by considering a double harmonic oscillator. 
This is much like a $0+1$ dimensional world instead of the usual $3+1$ 
dimensional one.
In section 6 we go into a study of the naive vacuum world.
In section 7 we illustrate how to rewrite an interaction into a
boundary conditions for the wave function by taking as an example the 
$\lambda(\varphi^+\varphi)^2$ theory.
Section 8 concerns some discussion of what analyticity properties of
singularities the wave function  is forced to have.
This study is motivated as a prerequisite for the CPT-like theorem
which is presented in section 9.
In section 10 a proof of the theorem is presented for the case of  a
$|\varphi|^4$ theory.
The relation to the usual CPT theorem is explained in section 11.
Finally in section 12 we give conclusions.


\section{The analytic harmonic oscillator}

In this section we consider as an exercise the formal problem of the
harmonic oscillator with the requirement of analyticity of the wave
function.
This will turn out to be crucial for our treatment of bosons with a
Dirac sea method analogous to the fermions.
In this exercise the usual requirement that the wave function
$\psi(x)$ should be square integrable

\begin{equation}
\int^\infty_{-\infty}|\psi(x)|^2dx<\infty
\end{equation}
is replace by the one that 
\begin{equation}
\psi(x) {\rm \ is \ analytic \ in \ C}
\end{equation}
where a possible essential singularity at $x=\infty$ is allowed.
In fact for this harmonic oscillator we shall prove the following
theorem :

1) The eigenvalue spectrum $E$ for the equation
\begin{equation}
\left(-\frac{\hbar^2}{2m}\frac{\partial^2}{\partial x^2}+
\frac{1}{2}m\omega^2x^2\right)\psi(x)=E\psi(x)
\end{equation}
is given by 
\begin{equation}
E=(n+\frac{1}{2})\hbar\omega \ \ \\ (n\varepsilon Z)
\end{equation} 
with {\it any integer} $n$.

2)The wave functions for $n=0,1,2,\ldots$ are the usual ones

\begin{equation}
\varphi_n(x)=A_n e^{-\frac{1}{2}(\beta x)^2}H_n(\beta x) \ \ \ .
\end{equation}

Here $\beta^2=\frac{m\omega}{\hbar}$ and $H_n(\beta x)$ the Hermite
polynomials of $\beta x$ while 
$A_n=\sqrt{\frac{\beta}{\pi^{\frac{1}{2}}2^nn!}}$ .
For $n=-1,-2,\ldots$ the eigenfunction is given by 

\begin{equation}
\varphi_n(x)=\varphi_{-n-1}(ix)=A_{-n-1}e^{\frac{1}{2}(\beta x)^2}
	H_{-n-1}(i\beta x) \ \ \ . 
\end{equation}

3)The inner product is defined as the natural one given by 

\begin{equation}
<\psi_1|\psi_2>=\int_\Gamma\psi_1(x^*)^*\psi_2(x)dx
\label{2.7}
\end{equation}
where the contour is taken to be the one along the
real axis from $x=-\infty$ to $x=\infty$.
The $\Gamma$ should be chosen so that the integrand should go down 
to zero at $x=\infty$, but there remains some
ambiguity in the choice.
However if one chooses the same $\Gamma$ so all the negative $n$
states, the norm squares of these states have an alternating sign.
In fact for the path $\Gamma$ along the imaginary axis from
$-i\infty$ to $i\infty$, we obtain 

\begin{eqnarray}
<\varphi_n|\varphi_m>&=&
	\int^{i\infty}_{-i\infty}\varphi_n(x^*)^*\varphi_n(x)dx\nonumber\\
&=&-(-1)^n
\label{2.8}
\end{eqnarray}

The above 1)-4) constitute the theorem.

Proof of this theorem is rather trivial.
We may start with consideration of large numerical $x$ behavior of a
solution to the eigenvalue equation.
Ansatze for the wave function is made in the form

\begin{equation}
\psi(x)=f(x)e^{\pm\frac{1}{2}(\beta x)^2}
\end{equation}
and we rewrite the eigenvalue equation as 

\begin{equation}
\frac{f''(x)}{\beta^2f(x)}\pm\frac{2f'(x)}{\beta f(x)}\beta x =
	-\frac{E\mp\frac{1}{2}\omega\hbar}{\omega\hbar}\ \ \ .
\label{2.10}
\end{equation}
If we use the approximation that the term $f''(x)/\beta^2f(x)$ is
dominated by the term $\pm\frac{2f'(x)}{\beta f(x)}\beta x$ for large
$|x|$, eq.(\ref{2.10}) reads

\begin{equation}
\frac{d\log f(x)}{d\log
x}=\frac{\mp E+\frac{1}{2}\omega\hbar}{\omega\hbar} \ \ \ .
\end{equation}
Here the right hand side is a constant $n$ which is yet to be shown to 
be an integer and we get as the large $x$ behavior

\begin{equation}
f(x)\sim x^n
\end{equation}
The reason that $n$ must be integer is that the function $x^n$ will
otherwise have a cut.
Thus requiring that $f(x)$ be analytic except for $x=0$ we must have

\begin{equation}
\mp E=-\frac{1}{2}\omega\hbar + n\hbar\omega
\end{equation}
For the upper sign the replacement $n\rightarrow -n-1$ is made and we can
always write 

\begin{equation}
E=\frac{1}{2}\hbar\omega+n\hbar\omega
\end{equation}
where $n$ takes not only the positive and zero integers
$n=0,1,2,\ldots$, but also the negative series $n=-1,-2 \ldots$ .

Indeed it is easily found that for negative $n$ the wave function is 

\begin{equation}
\varphi_n(x)=\varphi_{-n-1}(iX) = A_{-n-1}e^{\frac{1}{2}(\beta
x)^2}H_{-n-1}(i\beta x)
\end{equation}

Next we go to the discussion of the inner product which we define by
eq.(\ref{2.7}).
If the integrand $\psi_1(x^*)^*\psi_2(x)$ goes to zero as
$x\rightarrow\pm\infty$ the contour $\Gamma$ can be deformed as
usual. 
But when the integrand does not go to to zero, we may have to define inner
product by an analytic continuation of the wave functions from the
usual positive sector ones that satisfy
$\int|\psi(x)|^2dx<\infty$ .
If we choose $\Gamma$ to be the path along the imaginary axis from 
$x=-i\infty$ to $x=i\infty$, the inner product takes the form 

\begin{eqnarray}
<\varphi_n|\varphi_m>&=&
	\int^{i\infty}_{-i\infty}\varphi_n(x^*)^*\varphi_m(x)dx\nonumber\\
&=&i\int^\infty_{-\infty}
\varphi_{-n-1}\left(i\left(i\xi\right)^*\right)^*
	\varphi_{-m-1}\left(i\left(i\xi\right)\right)d\xi 
\label{2.16}
\end{eqnarray}
where $x$ along the imaginary axis is parameterized by $x=i\xi$ with a
real $\xi$ .
From eq.(\ref{2.16}) we obtain for the negative $n$ and $m$,

\begin{equation}
<\varphi_n|\varphi_m>=-i(-1)^m\delta_{nm}
\label{2.17}
\end{equation}
so that

\begin{equation}
\parallel\varphi_n\parallel^2=-i(-1)^n \ \ \ .
\label{2.18}
\end{equation}
We notice that the norm square has the alternating sign depending on
the even or odd negative $n$, when the contour $\Gamma$ is kept fixed. 

The reason why there is a factor $i$ in eq.(\ref{2.18}) can be understood as
follows :
when the complex conjugation for the definition of the inner product
(\ref{2.7}) is taken, the contour $\Gamma$ should also be complex
conjugated

\begin{equation}
<\psi_1|\psi_2>^*=\int_{\Gamma^*}\psi_1(x^*)\psi_2(x)^* dx
\end{equation}
Thus if $\Gamma$ is described by $x=x(\xi)$ as 

\begin{equation}
\Gamma=\{x(\xi)|-\infty<\xi<\infty \ \ \ : \ \ \ \xi ={\rm real}\}
\end{equation}
then $\Gamma^*$ is given by 

\begin{equation}
\Gamma^*=\{x^*(\xi)|-\infty<\xi<\infty\ \ \ ,\ \ \ \xi={\rm real}\} \ \ \ .
\end{equation}
So we find 

\begin{equation}
<\psi_1|\psi_2>^*=
	\int_{-\infty<\xi<\infty}\psi_2(x(\xi)^*)^*\psi_1(x(\xi))
	\frac{dx(\xi)^*}{dx(\xi)} dx(\xi)
\end{equation}
which deviates from $<\psi_2|\psi_1>$ by the factor
$dx(\xi)^*/dx(\xi)$ in the integrand.
In the case of $x(\xi)=i\xi \ , \ dx(\xi)^*/dx(\xi)=-1$ so that 

\begin{equation}
<\psi_1|\psi_2>^*=-<\psi_2|\psi_1>
\end{equation}
for the eigenfunctions of the negative sector.
From this relation the norm square is purely imaginary.

It may be a strange convention and  we may change the inner product 
eq.(\ref{2.7}) by a new one defined by 

\begin{equation}
<\psi_1|\psi_2>_{new}=\frac{1}{i}<\psi_1|\psi_2>
\end{equation}
so as to have the usual relation also in the negative sector

\begin{equation}
<\psi_1|\psi_2>^*_{new}=<\psi_2|\psi_1>_{new}
\end{equation}
Thus we conclude that the norm square is given by eq.(\ref{2.17}).


\section{The treatment of the Dirac sea for bosons}

We shall make use of the extended harmonic oscillator described in
previous section to quantize bosons.

As is well known in a non-relativistic theory a second quantized
system of bosons may be described by using an analogy with a system
of harmonic oscillators ;
one for each state in an orthonormal basis for the single particle.
The excitation number $n$ of the harmonic oscillator is identified
with the number of bosons present in that state in the basis to which
the oscillator corresponds.

For instance, if we have a system with $N$ bosons its state is
represented by the symmetrized wave function

\begin{equation}
\psi_{\alpha_1\ldots\alpha_N} \ ( \stackrel{\rightharpoonup}{x_1},
\ldots , \stackrel{\rightharpoonup}{x_N})
\end{equation}
where the indices $\alpha_1,\alpha_2\ldots,\alpha_N$ indicate the
intrinsic quantum numbers such as spin.
In a energy and momentum eigenstate
$k=(\stackrel{\rightharpoonup}{k},+)$ or 
$k=(\stackrel{\rightharpoonup}{k},-)$
we may write

\begin{eqnarray}
K_{\rm pos}&=&\{(\stackrel{\rightharpoonup}{k},+) |
\stackrel{\rightharpoonup}{k}\} \ , \\
K_{\rm neg}&=&\{(\stackrel{\rightharpoonup}{k},-) |
\stackrel{\rightharpoonup}{k}\} \  .
\end{eqnarray}
and $K=K_{\rm pos}\cup K_{\rm neg}$ .
We expand 
$\psi_{\alpha_1\ldots\alpha_N}
(\stackrel{\rightharpoonup}{x_1},\ldots,\stackrel{\rightharpoonup}{x_N})$ 
in terms of an orthonormal basis of single particle states
$\{\varphi_{k;\alpha} \ (\stackrel{\rightharpoonup}{x})\}$
with
$k\epsilon K$.
It reads

\begin{eqnarray}
|\psi>&=&\psi_{\alpha_1\ldots,\alpha_N}(\stackrel{\rightharpoonup}{x_1}, 
	\ldots,\stackrel{\rightharpoonup}{x_N})\nonumber\\
&=&\sum C_{{k_1},\ldots,k_N}\frac{1}{N!}\sum_{\rho\epsilon S_N}\nonumber\\
&&\varphi_{k_{\rho(1)}\alpha_1}
		(\stackrel{\rightharpoonup}{x_1})
	\varphi_{k_{\rho(2)}\alpha_2}
		(\stackrel{\rightharpoonup}{x_2})\cdots
	\varphi_{k_{\rho(N)}\alpha_N}
		(\stackrel{\rightharpoonup}{x_N})\ \ \ .
\end{eqnarray}
The corresponding state of the system of harmonic oscillators is given
by

\begin{equation}
|\psi > = \sum_{k_1,\ldots,k_N} C_{k_1,\ldots k_N}
	\prod_{k\epsilon K} | n_k=
	\sharp\{i | i=1,\cdots,N \ {\rm and} \ k_i=k\}>
\end{equation}
where $|n_k>$ represent the state of the $k$-th harmonic oscillator.

The harmonic oscillator is extended so as to have the negative $n_k$ 
values of the excitation number.
This corresponds to that the number of bosons $n_K$ in the single particle 
states could be negative.
In the non-relativistic case one can introduce the creation and
annihilation operators $a_k$ and $a^+_k$ respectively.
In the harmonic oscillator formalism these are the step operators for
the $k$th harmonic oscillator,

\begin{eqnarray}
a^+_k |n_k> &=&\sqrt{n_k+1} \ | n_k+1>\\
a_k|n_k > &=& \sqrt{n_k} \ | n_k-1> \ \ .
\end{eqnarray}

It is also possible to introduce creation and annihilation operators
for arbitrary states $|\psi > $

\begin{eqnarray}
a^+(\psi)&=&\sum_{k\epsilon K}<\varphi_k|\psi> a^+_k\\
a(\psi)&=&\sum_{k\epsilon K}a_k<\varphi_k|\psi> \ \ .
\end{eqnarray}
We then find 

\begin{eqnarray}
[a(\psi'),a^+(\psi)]&=&\sum_{k,k'}<\psi'|\varphi_{k'}>
	[a_{k'},a_k]<\varphi_k|\psi>\nonumber\\
&=&<\psi'|\psi> \ \ .
\end{eqnarray}
in which the right hand side contains an indefinite Hilbert product.
Thus if we perform this naive second quantization, the possible
negative norm square will be inherited into the second quantized
states in the Fock space.

Suppose that we had chosen the basis such that for some subset
$K_{\rm pos}$ the norm square is unity

\begin{equation}
<\varphi_k|\varphi_k>= 1 \ \ \  {\rm for} \ \  k\epsilon K_{\rm pos} 
\end{equation}
while for the complement set
$K_{\rm neg}=K\backslash K_{\rm pos}$ it is 

\begin{equation}
<\varphi_k|\varphi_k>=-1\ \ \ {\rm for} \ k\epsilon K_{\rm neg} \ \ .
\end{equation}
Thus any component of a Fock space state must have negative norm
square if it has an odd number of particles in states of
$K_{\rm neg}$.

In the analogy to the harmonic oscillator, we have the following
signs of the norm square in the naive second quantization 

\begin{equation}
<n_k|m_k> = \delta_{n_km_k}(-1)^{n_k}
\end{equation}
for $k\epsilon K_{\rm neg}$ where $n_k$ and $m_k$ denote the usual
nonzero levels. 
With use of our extended harmonic oscillators we end up with a system
of norm squared as follows :

\noindent
For $k\epsilon K_{\rm pos}$ 

\begin{equation}
\begin{array}{ll}
&<n_k|m_k> \\
&=\left\{
\begin{array}{ll}
\delta_{n_km_k}&{\rm for} \ n_k,m_k=0,1,2,\ldots\\
i\delta_{nkm_k}(-1)^{n_k}& {\rm for} \ n_k,m_k=-1,-2\ldots\\
\infty&{\rm for} \ n_k \ {\rm and} \ m_k \ {\rm in \ different \ sectors}\  .
\end{array}
\right.
\end{array}
\end{equation}

\noindent
For $k\epsilon K_{\rm neg}$ 
\begin{equation}
\begin{array}{ll}
&<n_k|m_k> \\
&=\left\{
\begin{array}{ll}
\delta_{n_km_k}(-1)^{n_k}&{\rm for} \ n_k,m_k=0,1,2,\ldots\\
i\delta_{n_km_k}& {\rm for} \ n_k,m_k=-1,-2\ldots\\
\infty&{\rm for} \ n_k \ {\rm and} \ m_k \ {\rm in \ different \ sectors}\  .
\end{array}
\right.
\end{array}
\end{equation}
We should bear in mind that the trouble of negative norm square is
solved by putting minus one particle into each state with $k\epsilon
K_{\rm neg}$.
Thereby we get it restricted to negative number of particles in these
states.
Thus we have to use the inner product 
$<n_k|m_k>=i\delta_{n_km_k}$, which makes the Fock space sector be a
good positive definite Hilbert space apart from the overall factor
$i$.

We may formulate our procedure in the following. 
The naive vacuum may be described by the state in terms of the ones of 
the harmonic oscillators as

\begin{equation}
| \,{\rm naive \ vac.} > =
	 \prod_{k\epsilon K} | 0>_{\stackrel{kth}{osc}} \ \ \ .
\end{equation}
On the other hand the correct vacuum is given by the state 

\begin{equation}
| \,{\rm correct \ vac.} > =
	\prod_{k\epsilon K_{\rm pos}}|0>_{\stackrel{kth}{osc}} \cdot
	\prod_{k\epsilon K_{\rm neg}} | -1 >_{\stackrel{kth}{osc}}
\end{equation}
where the states $|-1>$ in $K_{\rm neg}$ are the ones with minus one
particles.

We proceed to the case of relativistic integer spin particles of which 
inner product is indefinite by Lorentz invariance

\begin{equation}
\int\psi^\ast(\stackrel{\rightharpoonup}{x},t) 
	\stackrel{\leftrightarrow}{\partial_t}
	\psi(\stackrel{\rightharpoonup}{x},t) d^3
	\stackrel{\rightharpoonup}{x} \ \ \ .
\end{equation}
The energy of the naive vacuum is given by

\begin{equation}
{\rm E_{naive \ vac.}}=
	\sum_{k\epsilon K}\frac{1}{2}\omega_k = 0 \ \ \ .
\end{equation}
By adding minus one particle to each negative energy state 
$\varphi_{k-}$ with $k\epsilon K_{\rm neg}$ the second quantized system 
is brought into such a sector that it is in the ground state, that is
the correct vacuum.
The energy of it is given by 

\begin{eqnarray}
{\rm E_{correct \ vac.}}&=&
	\sum_{k\epsilon K}\frac{1}{2}\omega_k-
	\sum_{k\epsilon K_{\rm neg}}\frac{1}{2}\omega_k\\
&=&\sum_{k\epsilon K}\frac{1}{2}|\omega_k | =
	\sum_{k\epsilon K_{\rm pos}}\omega_k \ \ \ .
\end{eqnarray}
It should be stressed that only inside the sector we obtain the ground
state in this way.
In fact with the single particle negative energies for bosons, the
total hamiltonian may have no bottom.
So if we do not add minus one particle to each single particle negative
energy state, one may find a series of states of which energy goes to 
$-\infty$.
However, by adding minus one particle we get a state of the second
quantized system in which there is the barrier due to the laser
effect.
This barrier keeps the system from falling back to lower energies as
long as polynomial interaction in $a^+_k$ and $a_k$ are concerned.

In the above calculation for the relativistic case we have

\begin{equation}
{\rm E_{correct \ vac}> E_{naive \ vac}} \ \ \ .
\end{equation}
Thus at the first sight the correct vacuum looks unstable.
However which vacuum has lower energy is not important for the
stability of a certain proposal of vacuum.
Rather the range of allowed energies for the sector of the vacuum
proposal is important.
To this end we define the energy range ${\rm E_{range}}$ of the vacuum
by

\begin{equation}
{\rm E_{range}}(|{\rm vac}>)=\{{\rm E}\}
\end{equation}
where ${\rm E}$ denotes an energy in a state which can be reached from 
$|{\rm vac}>$ by some operators polynomial in $a^+$ and $a$.
Thus for naive vacuum

\begin{equation}
{\rm E_{range}(|naive \ vac>)}=(-\infty,\infty)
\end{equation}
while for correct vacuum

\begin{equation}
{\rm E_{range}(|correct \ vac>)}=
	[\sum_{k\epsilon K}\frac{1}{2}|\omega_K|, \infty] \ \ \ .
\end{equation}
Once the vacuum is brought into the correct vacuum state, it is no
longer possible to add particles to the state with $K_{neg}$, due to
the barrier rather to subtract particles.
Thus $a_k$ with $k\epsilon K_{\rm neg}$ can act on
$|-1>_{\stackrel{kth}{osc}}$ with arbitrary number of times as

\begin{equation}
(a_k)^n |-1>_{\stackrel{kth}{osc}}=
	\sqrt{|n|!} \ | -1-n >_{\stackrel{kth}{osc}} \ \ \ .
\end{equation}
These subtractions we may call holes which correspond to addition of
antiparticles.

It is natural to switch notations from dagger to non dagger one by
defining 

\begin{equation}
b^+(-\stackrel{\rightharpoonup}{k},{\rm anti})=
	a(\stackrel{\rightharpoonup}{k}, -) 
\end{equation}
and vice versa 
where $k=(\stackrel{\rightharpoonup}{k}, -)$ is a $\omega<0$ state
with 3-momentum $\stackrel{\rightharpoonup}{k}$.
The operator $b^+(-\stackrel{\rightharpoonup}{k}, {\rm anti})$
denotes an creation of antiparticle with momentum
$\stackrel{\rightharpoonup}{k}$ and positive energy $-\omega>0$ .
This is exactly the usual way of treatment of the second quantization
for bosons.
The commutator of these operators reads
\begin{equation}
[b(\stackrel{\rightharpoonup}{k}, {\rm anti}),
	b^+(\stackrel{\rightharpoonup}{k'},{\rm anti})]=
	\delta_{\stackrel{\rightharpoonup}{k}
		\stackrel{\rightharpoonup}{k'}} \ \ \ .
\end{equation}
It should be noticed that in the boson case the antiparticles are also 
holes.
Before closing this section two important issues are discussed.
The first issue is that there are four vacua in our approach of
quantization.

We have argued that we can obtain the correct vacuum by modifying the
naive vacuum so that one fermion is filled and one boson removed from 
each single particle negative energy state.
This opens the possibility of considering naive vacuum and associated
world of states where there exists a few extra particles. 
The naive vacuum should be considered as a playground for study of the 
correct vacuum.
It should be mentioned that once we start with one of the vacua and
work by filling the negative energy states or removing from it, we may 
also do so for positive energy states. 
In this way we can think of four different vacua which are
illustrated symbolically as type a-d in Fig.1.

\begin{figure}[h]
\epsfysize=120mm
\epsfbox{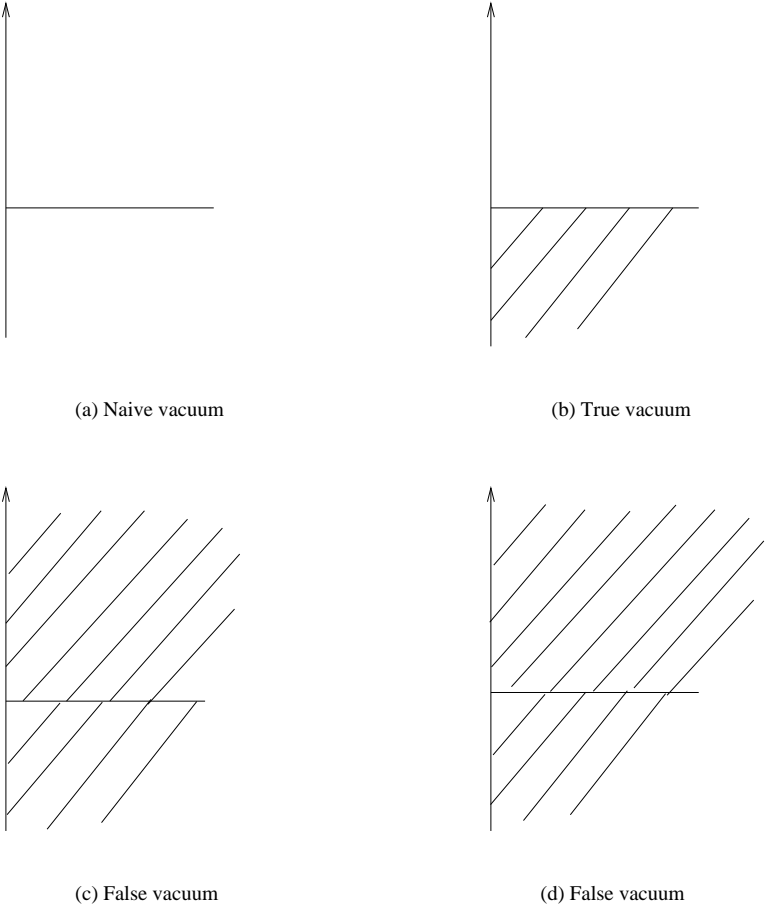}
\caption{Four types of vacua}{
There are four possible types of vaua for bosons as well as fermions.
In Figures (a) - (d) the shaded states denote that they are all filled 
by one particle for fermions and minus one particle for bosons. 
The unshaded states are empty.
}
\end{figure}

As an example consider the type c vacuum.
In this vacuum the positive energy states are modified by filling the
positive energy states by one fermion but removing from it by one
boson, while the negative energy states are not modified.
Thus the single particle energy spectrum has a top but no bottom.
Inversion of the convention for the energy would not be possible to
be distinguished by experiment as far as free system is concerned.
However it would have negative norm square for all bosons and the
interactions would work in an opposite manner. 
We shall show in the later sections that there exists a trick of
analytic continuation of the wave function to circumvent this
inversion of the interaction.

Another issue to be mentioned is the CPT operation on those four
vacua. 
The CPT operation on the naive vacuum depicted as type a vacuum in
Fig.1 does not get it back.
The reason is that by the charge conjugation operator C all the holes
in the negative energy states are, from the correct vacuum point
of view, replaced by particles of corresponding positive energy
states.
Thus acting CPT operator on the naive vacuum sends into the type c
vacuum because the positive energy states is modified while the
negative ones remain the same.
This fact may be stated that in the naive vacuum CPT symmetry is
spontaneously broken.

However in section 9 we shall put forward another CPT-like theorem in
which the CPT-like symmetry is preserved in the naive vacuum but
broken in the correct one.

Before going into presenting the new CPT-like theorem in Section 9 
we mention some properties of the world around the naive vacuum 
where there is only a few particles. 
The terminology of the world around a vacuum is used for the one with 
a superposition of such a states that it deviates from the vacuum in
questions by a finite number of particles and that the boson does not 
cross the barrier. 
Since the naive vacuum has no particles and we can add positive numbers 
of particles which, however, can have both positive and negative
energies.
The correct vacuum may similarly have a finite number of particles and 
holes in addition to the negative energy seas.

\section{Wave functional formulation}

In this section we develop the wave functional formulation of field
theory in the naive vacuum world.

When going over field theory corresponding to the naive field
quantization

\begin{eqnarray}
\varphi(\stackrel{\rightharpoonup}{x},t)&=&
	\sum_{\stackrel{\rightharpoonup}{p},{\rm sig}n}
	\frac{1}{\sqrt{|\omega |}}
	a(\stackrel{\rightharpoonup}{p},{\rm sig}n)
	e^{-i\omega t+i\stackrel{\rightharpoonup}{p}\cdot
		\stackrel{\rightharpoonup}{x}}
\\
\pi(\stackrel{\rightharpoonup}{x},t)&=&
	\sum_{\stackrel{\rightharpoonup}{p},{\rm sig}n}
	\frac{1}{\sqrt{|\omega |}}
	a(\stackrel{\rightharpoonup}{p},{\rm sig}n)
	\cdot({\rm sig}n)	
	e^{-i\omega t+i\stackrel{\rightharpoonup}{p}\cdot
		\stackrel{\rightharpoonup}{x}}
\end{eqnarray}
we have a wave functional $\Psi[\varphi]$.
For each eigenmode 
$\omega\varphi_{\stackrel{\rightharpoonup}{p}}+
	i\pi_{\stackrel{\rightharpoonup}{p}}$ ,
where $\varphi_{\stackrel{\rightharpoonup}{p}}$ are the 3-spatial
Fourier transforms of $\varphi(x)$ and
$\pi_{\stackrel{\rightharpoonup}{p}}$ is conjugate momentum, we have 
an extended harmonic oscillator described in section 2.
In order to see how to put the naive vacuum world into a wave
functional formulation, we investigate the Hamiltonian and the
boundary conditions for a single particle states with a general norm
square.

Let us imagine that we make the convention in which the $n$-particle
state be

\begin{equation}
A_nH_n(x)
\end{equation}
with $H_n$ the Hermite polynomial.
Thus

\begin{equation}
|n>=A_nH_n(x)|0> \ \ \ .
\end{equation}
On the other hand $n$-excited state in the harmonic oscillator is given
by 

\begin{equation}
A_nH_n(x)\beta e^{-\frac{1}{2}(\beta x)^2}
\end{equation}
with $\beta^2=\frac{\hbar}{m\omega}$.
We can vary the normalization while keeping the convention as 
\begin{equation}
<n|n>=\beta^{-2n}<0|0> \ \ \ .
\end{equation}
We can consider $\beta^{-2}$ as the norm square of the single particle 
state corresponding to the harmonic oscillator.

Now the Hamiltonian of the harmonic oscillator is expressed in terms
of 
$\omega$ and $\beta^{-2}$ as

\begin{equation}
H=-\frac{\omega}{2<s.p|s.p>}\frac{d^2}{dx^2}+\frac{1}{2}<s.p|s.p>
\omega x^2
\label{4.7}
\end{equation}
where $|s.p>$ denotes the single particle state and 

\begin{equation}
<s.p|s.p>=m\omega
\end{equation}
with $\hbar=1$.
Remark that if one wants $<s.p|s.p>$ negative for negative $\omega$
one should put the usual signs on the coefficient in $H$.
However in this case $\beta$ turns out to be pure imaginary and
$e^{-\frac{1}{2}(\beta X)^2}$ blows up so that the wave functions
become like the one in the extended negative sector discussed before.

By passing to the correct vacuum world by removing one particle from
each negative energy state, the boundary conditions for the wave
functional are changed so as to converge along the real axis for all
the modes.
In fact they were along the imaginary axis for the negative energy
modes in the naive vacuum.

From the fact that the form of the Hamiltonian in the wave functional
formalism must be the same as for the correct vacuum we can easily
write down the Hamiltonian.
For instance using the conjugate variable $\pi$

\begin{equation}
\pi(\stackrel{\rightharpoonup}{x})=-i
	\frac{\delta}{\delta\varphi(\stackrel{\rightharpoonup}{x})}
\end{equation}
the free Hamiltonian becomes 

\begin{equation}
H_{\rm free}=\int\frac{1}{2}\left\{|\pi(\stackrel{\rightharpoonup}{x})|^2+
	|\bigtriangledown\varphi(\stackrel{\rightharpoonup}{x})|^2+
	m|\varphi(\stackrel{\rightharpoonup}{x})|^2\right\}
	d^3\stackrel{\rightharpoonup}{x} \ \ \ .
\end{equation}
This acts on the wave functional 

\begin{eqnarray}
&&H_{\rm free}\Psi[\varphi]\nonumber\\
&&=\frac{1}{2}\int\left\{-
	\frac{\delta^2}{\delta\varphi(\stackrel{\rightharpoonup}{x})^2}+
	|\bigtriangledown\varphi(\stackrel{\rightharpoonup}{x})|^2+
	m^2|\varphi(\stackrel{\rightharpoonup}{x})|^2\right\}
	\Psi[\varphi]\ \ \ .
\end{eqnarray}
The inner product for the functional integral is given by

\begin{eqnarray}
<\Psi_1|\Psi_2>&=&
	\int\Psi_1[(Re\varphi)^\ast,
		(Im\varphi)^\ast]^\ast\nonumber\\
&&\cdot\Psi_2[Re\varphi,Im\varphi]
	{\cal D}Re\varphi\cdot{\cal D}Im\varphi \ \ \ ,\nonumber
\end{eqnarray}
where the independent functions are
$Re\varphi(\stackrel{\rightharpoonup}{x})$ and 
$Im\varphi(\stackrel{\rightharpoonup}{x})$ .
In order to describe the wave functional theory of the naive vacuum
world we shall make a formulation in terms of the convergence
condition along the real function space for 
$Re\varphi$ and $Im\varphi$ .
In fact we go to the representation in which 
$\Psi [Re\varphi, Im\pi]$ is expressed by means of $\varphi$ and
$\pi$ representations.

We would like to organize so that the boundary conditions for the
quantity 
$\omega\varphi_{\stackrel{\rightharpoonup}{k}}+
i\pi_{\stackrel{\rightharpoonup}{k}}$
are convergent in the real axis for $\omega>0$ while for $\omega<0$  
they are so in the imaginary axis.
We may think the real and imaginary parts of
$(\omega\varphi_{\stackrel{\rightharpoonup}{k}}+
i\pi_{\stackrel{\rightharpoonup}{k}})$ 
separately.
Then the requirement of convergence in the correct vacuum should be
that for $\omega<0$ the formal expression

\begin{eqnarray}
Re(\omega\varphi_{\stackrel{\rightharpoonup}{k}}+
i\pi_{\stackrel{\rightharpoonup}{k}})&=&
	\frac{\omega}{2}\left\{(Re\varphi)_{\stackrel{\rightharpoonup}{k}}+
		(Re\varphi)_{-\stackrel{\rightharpoonup}{k}}\right\}
		\nonumber\\
&&-\frac{1}{2}\left\{(Im\pi)_{\stackrel{\rightharpoonup}{k}}+
		(Im\pi)_{-\stackrel{\rightharpoonup}{k}}\right\}
\end{eqnarray}	 
and

\begin{eqnarray}
Im(\omega\varphi_{\stackrel{\rightharpoonup}{k}}+
i\pi_{\stackrel{\rightharpoonup}{k}})&=&
	\frac{\omega}{2}\left\{(Im\varphi)_{\stackrel{\rightharpoonup}{k}}+
		(Im\varphi)_{-\stackrel{\rightharpoonup}{k}}\right\}
		\nonumber\\
&&+\frac{1}{2}\left\{(Re\pi)_{\stackrel{\rightharpoonup}{k}}+
		(Re\pi)_{-\stackrel{\rightharpoonup}{k}}\right\}
\end{eqnarray}
are purely imaginary along the integration path for which the
convergence is required.

We may use the following parameterization in terms of the two real
functions $\chi_1$ and $\chi_2$ :

\begin{eqnarray}
Re\varphi&=&-(1+i)\chi_1-(1-i)\chi_2\nonumber\\
Im\pi&=&(1-i)\chi_1+(1+i)\chi_2\nonumber
\end{eqnarray}
By this parameterization the phases of
$\omega\varphi_{\stackrel{\rightharpoonup}{k}}+
	i\pi_{\stackrel{\rightharpoonup}{k}}$ 
lay in the intervals 

\begin{eqnarray}
&&]\frac{\pi}{4} \ , \ \frac{3\pi}{4}[ \ \ \ 
	{\rm for} \ \ \  \omega<0\nonumber\\
{\rm and }&&\nonumber\\
&&]-\frac{\pi}{4} \ , \ \frac{\pi}{4}[ \ \ \ 
	{\rm for} \ \ \  \omega>0\nonumber 
\end{eqnarray}
modulo $\pi$.
They provide the boundary condition for the naive vacuum world when
convergence of ${\cal D}\chi_1{\cal D}\chi_2$ integration is required.

In this way  we find the naive vacuum world with usual wave functional 
hamiltonian operator.
However we do not require the usual convergence condition 

\begin{equation}
\int\Psi(Re\varphi, Im\varphi)^\ast
\Psi(Re\varphi, Im\varphi){\cal D}Re\varphi{\cal D}Im\varphi < \infty
\label{4.14}
\end{equation}
but instead require 

\begin{equation}
<\Psi|\Psi>=\int\Psi[(Re\varphi)^\ast,(Im\pi)^\ast]^\ast
	\Psi[Re\varphi, Im\varphi]{\cal D}\chi_1{\cal D}\chi_2 <
\infty
\end{equation}
where the left hand side is defined along the path with
$\chi$-parameterization. 
The inner product corresponding to this functional contour is 

\begin{eqnarray}
&<\Psi_1|\Psi_2>&=
	\int\Psi_1[-(1-i)\chi_1-(1+i)\chi_2,(1+i)\chi_1+(1-i)\chi_2]^\ast
	\nonumber\\
&&\Psi_2[-(1+i)\chi_1-(1-i)\chi_2,(1-i)\chi_1+(1+i)\chi_2]
	{\cal D}\chi_1{\cal D}\chi_2 \ \ \ . 
\label{4.16}
\end{eqnarray}
This is not positive definite, and that is related to the fact that there
are lots of negative norm square states in the Fock space in the naive
vacuum world.

The method of filling the vacuum for fermions now becomes for the case 
of bosons that first in the naive vacuum we have the strange convergence
condition eq.(\ref{4.14}).
We then go to the correct vacuum by switching the boundary conditions
to the convergence along real axis e.g. $Re\varphi$ and $Im\pi$ real.


\section{Double harmonic oscillator}

To illustrate how our functional formalism works we consider as a simple
example a double harmonic oscillator.
It is relevant for us in the following three points:

\begin{enumerate}
\renewcommand{\labelenumi}{\arabic{enumi})}
\item It is the subsystem of a field theory which consists of two single 
particle states with 
$p^\mu=
(\stackrel{\rightharpoonup}{p},\omega(\stackrel{\rightharpoonup}{p}))$
and 
$-p^\mu=
(-\stackrel{\rightharpoonup}{p},-\omega(\stackrel{\rightharpoonup}{p}))$
for $\omega(\stackrel{\rightharpoonup}{p})>0$ .
\item It could correspond a single 3-position field where 
the gradient interaction is ignored.
\item It is a $0+1$ dimensional model field theory.
\end{enumerate}

We start by describing the spectrum for free case corresponding to a
two state system in which
the two states have opposite $\omega$'s.
The boundary conditions in the naive vacuum world is given by 

\begin{eqnarray}
&&\int
\psi\left(\left(Re\varphi\right)^\ast , \left(Im\Pi\right)^\ast\right)^\ast
\psi\left(Re\varphi , Im\Pi\right)d\chi_1\chi_2
\nonumber\\
&&=\int
\psi\left(-\left(1-i\right)\chi_1-\left(1+i\right)\chi_2 ,
	\left(1+i\right)\chi_1+\left(1-i\right)\chi_2\right)^\ast
	\nonumber\\
&&\cdot \ \psi\left(-\left(1+i\right)\chi_1-\left(1-i\right)\chi_2 ,
	\left(1-i\right)\chi_1+\left(1+i\right)\chi_2\right) 
	d\chi_1d\chi_2 < \infty
\label{5.1}
\end{eqnarray}
which is similar to eq.(\ref{4.16}).
However in eq.(\ref{5.1})
the quantities $\chi_1$ and $\chi_2$ are not functions, but just real
variables.
Here we use a mixed representation in terms of position variables
$Re\varphi$ and $Im\varphi$ and conjugate momenta 

\begin{equation}
Re\pi=-i\frac{\partial}{\partial Re\varphi} \ ,\ 
Im\pi=-i\frac{\partial}{\partial Im\varphi} 
\end{equation}

The Hamiltonian may be given by a rotationally symmetric 2 dimensional
oscillator because the two $\omega$'s are just opposite.
From eq.(\ref{4.7}) the coefficient of 
$\frac{\partial^2}{\partial Im\varphi^2}$ is 

\begin{equation}
\frac{-\omega}{2<s.p|s.p.>}
\end{equation}

and that of $(Im\varphi)^2$ is 

\begin{equation}
\frac{1}{2}<s.p.|s.p.>\omega
\end{equation}
where $|s.p>$ denotes the single particle state.
These  coefficients are the same for both oscillators and 
thus the Hamiltonian reads

\begin{eqnarray}
H&=&-\frac{1}{2}|\omega|\frac{\partial}{\partial\varphi}
	\frac{\partial}{\partial\varphi^\ast}+
	\frac{1}{2}|\omega|\varphi^\ast\varphi\nonumber\\
&=&\frac{1}{2}|\omega|\left(-
	\frac{\partial^2}{\partial Re\varphi^2}-
	\frac{\partial^2}{\partial Im\varphi^2}+
	Re\varphi^2+Im\varphi^2\right) \ \ \ .\nonumber
\end{eqnarray}
which is expressed in the mixed representation as 

\begin{equation}
H=\frac{1}{2}|\omega|\left(-\frac{\partial^2}{\partial Re\varphi^2}+
	Re\varphi+Im\pi^2-
	\frac{\partial^2}{\partial Im\pi^2}\right) \ \ \ .
\end{equation}
We may express $H$ in terms of the real parameterization $\chi_1$ and
$\chi_2$ by using the relations 

\begin{eqnarray}
Re\varphi&=&-(1+i)\chi_1-(1-i)\chi_2\nonumber \\
Im\pi&=&(1-i)\chi_1+(1+i)\chi_2 \ \ \ .\nonumber
\end{eqnarray}
It may be convenient to define

\begin{equation}
\chi_\pm=\sqrt{2}(\chi_2\pm\chi_1)
\end{equation}
so that 

\begin{equation}
H=\frac{1}{2}|\omega|\left(
	\frac{\partial^2}{\partial\chi^2_-}-\chi^2_- -
	\frac{\partial^2}{\partial\chi^2_+}+\chi^2_+\right) \ \ \ .
\end{equation}
The inner product takes the form

\begin{equation}
<\tilde{\psi_1}|\tilde{\psi_2}>=
	\int\tilde{\psi_1}(-\chi_-,\chi_+)^\ast
	\tilde{\psi_2}(\chi_-,\chi_+)d\chi_-d\chi_+
\end{equation}
where

\begin{equation}
\tilde{\psi_i}(\chi_-,\chi_-)=
\psi_i(-\sqrt{2}\chi_++i\sqrt{2}\chi_- , \sqrt{2}\chi_++i\sqrt{2}\chi_-)
\end{equation}

As to be expected the Hamiltonian turns out to be two uncoupled
harmonic oscillators expressed in terms of $\chi_-$and $\chi_+$.
The $\chi_+$ oscillator is usual one while $\chi_-$ one has the
following two deviations :
one is that it has over all negative sign.
The other one is that in the definition of the inner product $-\chi_-$
instead of $\chi_-$ is used in the bra-wave function.
This is equivalent to abandoning a parity operation
$\chi_-\rightarrow -\chi_-$ in the inner product.

The energy spectrum is made up from all combinations of a positive
contribution $|\omega|(n_++\frac{1}{2})$ with a negative one
$-|\omega|(n_-+\frac{1}{2})$ so that

\begin{equation}
E=|\omega|(n_+-n_-) \ \ \ .
\label{5.10}
\end{equation}
The norm square of these combination of the eigenstates are
$(-1)^{n_-{-1}}$
which equals to the parity under the $\chi_-$ parity operation 
$\chi_-\rightarrow -\chi_-$  .

If we consider the single particle state, the charge or the number of
particles is given by 

\begin{eqnarray}
Q&=&\frac{i}{4}\left\{
	\pi^+,\varphi\right\}-\frac{i}{4}\left\{\varphi^+, \pi\right\}
	\nonumber\\
&=&\frac{1}{2}\chi^2_+-\frac{1}{2}\frac{\partial^2}{\partial\chi^2_+}+
	\frac{1}{2}\chi^2_--\frac{1}{2}\frac{\partial^2}{\partial\chi^2_-}
	-1 \ \ \ .
\end{eqnarray}
This is simply a sum of two harmonic oscillator Hamiltonians with the
same unit frequency.
Thus the eigenvalue $Q'$ of $Q$ can only take positive integer or
zero.
For a given value $Q'$ that is number of particle in either of the
two states, the energy can vary from
$E=-|\omega|Q'$ to $E=|\omega|Q'$ in integer steps in $2|\omega|$ .
Thus we can put 

\begin{equation}
n_-=Q', Q'-1, Q'-2, \cdots, 0
\end{equation}
in 
the negative $\omega$ states and in the positive energy state we have 

\begin{equation}
n_+=Q-n_- \  .
\end{equation}
So the energy eq.(\ref{5.10}) can be written as 

\begin{equation}
E=|\omega|(Q-2n_-)
\label{5.14}
\end{equation}
which is illustrated in Fig.2(a).
By going to the convergence condition along the real axis we get the
usual theory with correct vacuum, see Fig.2(b).

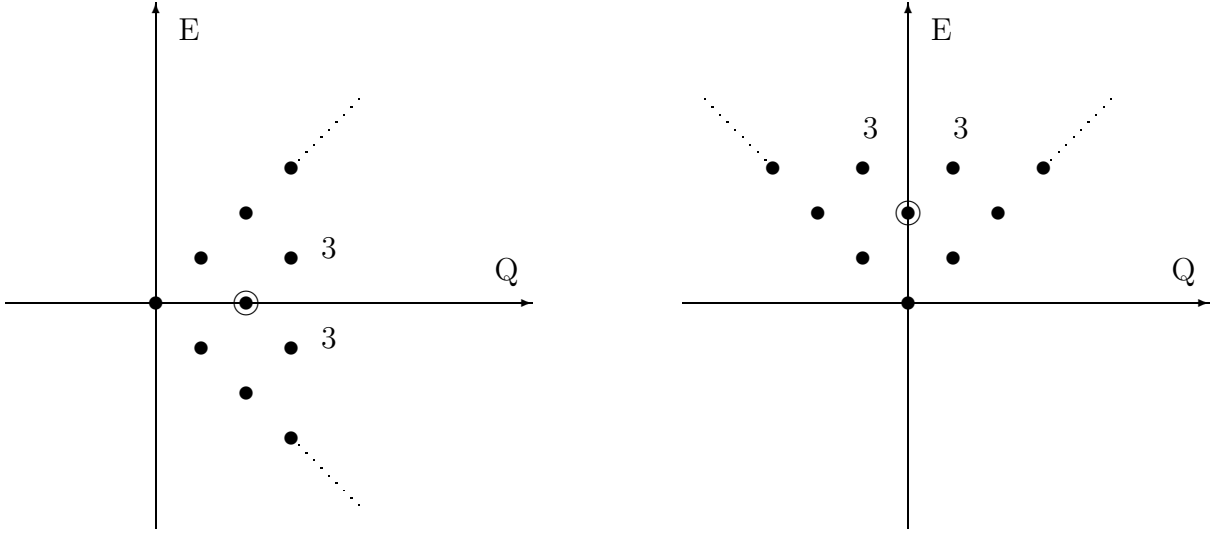
\begin{figure}[h]
\unitlength 1mm
\begin{picture}(150, 50)
\put(0,-30){\vector(0,1){70}}
\put(-20,0){\vector(1,0){70}}
\put(3,35){E}
\put(45,3){Q}
\put(0,0){\circle*{1.8}}
\put(6,6){\circle*{1.8}}
\put(12,12){\circle*{1.8}}
\put(18,18){\circle*{1.8}}
\put(6,-6){\circle*{1.8}}
\put(12,-12){\circle*{1.8}}
\put(18,-18){\circle*{1.8}}
\put(12,0){\circle{3}}
\put(12,0){\circle*{1.8}}
\put(18,6){\circle*{1.8}}
\put(18,-6){\circle*{1.8}}
\multiput(18,18)(1,1){10}{\line(0,0){0.1}}
\multiput(18,-18)(1,-1){10}{\line(0,0){0.1}}
\put(22,6){3}
\put(22,-6){3}
\put(100,-30){\vector(0,1){70}}
\put(70,0){\vector(1,0){70}}
\put(103,35){E}
\put(135,3){Q}
\put(100,0){\circle*{1.8}}
\put(106,6){\circle*{1.8}}
\put(112,12){\circle*{1.8}}
\put(118,18){\circle*{1.8}}
\put(94,6){\circle*{1.8}}
\put(88,12){\circle*{1.8}}
\put(82,18){\circle*{1.8}}
\put(100,12){\circle{3}}
\put(100,12){\circle*{1.8}}
\put(106,18){\circle*{1.8}}
\put(94,18){\circle*{1.8}}
\multiput(118,18)(1,1){10}{\line(0,0){0.1}}
\multiput(82,18)(-1,1){10}{\line(0,0){0.1}}
\put(106,22){3}
\put(94,22){3}
\end{picture}
\vskip40mm
(a) Naive vacuum
\hskip60mm
(b) True vacuum
\caption{charge vs energy in two state system }
{Naive vacuum case is depicted in Figure (a), while Figure (b) is for
the case of true vacuum.}
\end{figure}

The wave function of the naive vacuum is given by 

\begin{equation}
\psi_{n.\upsilon.}=
	N\exp\left(-\frac{1}{2}\chi^2_--\frac{1}{2}\chi^2_+\right)
\end{equation}
with a normalization constant $N$.
We may transform eq.(\ref{5.14}) in the mixed transformation back to the
position representation by Fourier transformation 

\begin{eqnarray}
\psi_{n.\upsilon.}(Re\varphi, Im\varphi)&=&
	\int e^{iIm\Pi\cdot Im\varphi}\psi_{n.\upsilon}
	(Re\varphi, Im\Pi)dIm\Pi\nonumber\\
&=&N\int e^{iIm\Pi\cdot Im\varphi}e^{Im\Pi\cdot Re\varphi}dIm\Pi \ \ .
	\nonumber\\
&=&N\delta(Im\varphi-iRe\varphi) \ \ \ .
\end{eqnarray}
Here $\delta$-function is considered as a functional linear in test
functions that is analytic and goes down faster than any power in
real directions and no faster than a certain exponential in imaginary
direction. 
This function may be called the distribution class $Z'$ according to
Gel'fand and Shilov \cite{gelfand}.
Thus our naive vacuum wave function is $\delta$-function that belongs
to $Z'$.

By acting with polynomials in creation and annihilation operators of
the naive vacuum state we obtain the expression of the form 

\begin{equation}
\sum_{n,m=0,1,\ldots}a_{n,m}(Re\varphi-iIm\varphi)^n
\delta^{(m)}(Re\varphi+Im\varphi) \ \ \ .
\label{5.17}
\end{equation}
Thus the wave functions of the double harmonic oscillator in the naive 
vacuum world are composed of eq.(\ref{5.17}).

As long as the charge $Q$ is kept conserved, even an interaction term
such as an anharmonic double oscillator with phase rotation symmetry, 
only the states of the form eq.(\ref{5.17})
can mix each other.
For such a finite quantum number $Q$ there is only a finite number of
these states of the form.
Therefore even the anharmonic oscillator would be reduced to finite
matrix diagonalization.
In this sense the naive vacuum world is more easier to solve
than the correct vacuum world.

In higher dimensions we may extend our result of the double harmonic
oscillator an obtain a $\delta$-functional wave functional for the
naive vacuum.
The naive vacuum world would involve polynomials in the combinations
that are not present in the $\delta$-functionals and their
derivatives.

\section{The Naive vacuum world}

In this section we shall investigate properties of the naive vacuum
world.
It is obvious that this world has the following five wrong properties
from the point of view of phenomenological applications:

\begin{enumerate}
\renewcommand{\labelenumi}{\arabic{enumi})}
\item There is no bottom in the energy.

\item The Hilbert space is not a true Hilbert space because it is 
not positive definite.
The states with an odd number of negative energy bosons get an extra
minus sign in the norm square.

We may introduce boundary conditions to make a model complete which
may be different for negative energy states.
As will be shown in section 8 in order to make an elegant CPT-like
symmetry we shall propose to take the boundary condition for the
negative energy states such that bound state wave functions blow up.

\item We cannot incorporate particles which are their own
antiparticles.
Thus we should think that all particles have some charges.

\item The naive vacuum world can be viewed as a quantum mechanical
system rather than second quantized field theory.
It is so because we think of a finite number of particles and the second
quantized naive vacuum world is in a superposition of various finite 
numbers of particles.

\item As long as we accept the negative norm square there is no
reason for quantizing integer spin particles as bosons and half
integer ones as fermions.
Indeed we may find the various possibilities as is shown in table 1.
In this table we recognize that the wellknown spin-statistics theorem
is valid only under the requirement that the Hilbert space is positive 
definite.
It should be noticed that in the naive vacuum world with integer spin
states negative norm squares exist anyway and so there is no
spin-statistics theorem.
When we go to the correct vacuum it becomes possible to avoid
negative norm square.
Then this calamity of indefinite Hilbert space is indeed avoided by
choosing the Bose or Fermi statistics according to the
spin-statistics theorem which is depicted in table 2.

\end{enumerate}

\begin{table}[h]
\begin{center}
\begin{tabular}{|c|c|c|}
\hline
\backslashbox{statistics}{spin}&
	$S=\frac{1}{2},\frac{3}{2},\ldots$&
	$S=0,1,\ldots$\\
\hline
Fermi-Dirac&$\|\cdots\|^2\geq 0$& Indefinite\\
\hline
Bose-Einstein&$\|\cdots\|^2\geq 0$ &Indefinite\\
\hline
\end{tabular}
\end{center}
\caption{Spin-statistics theorem for naive vacuum}
\end{table}

\vspace{10mm}
\begin{table}[h]
\begin{center}
\begin{tabular}{|c|c|c|}
\hline
\backslashbox{statistics}{spin}&
	$S=\frac{1}{2},\frac{3}{2},\ldots$&
	$S=0,1,\ldots$\\
\hline
Fermi-Dirac&$\|\cdots\|^2\geq 0$& Indefinite\\
\hline
Bose-Einstein&Indefinite&$\|\cdots\|^2\geq 0$ \\
\hline
\end{tabular}
\end{center}
\caption{Spin-statistics theorem for true vacuum}
\end{table}


\section{The $(\varphi^+\varphi)^2$ theory and boundary condition}

In this section we illustrate how to rewrite an interaction into a
boundary condition for the wave function by taking as an example the
scattering in the $\lambda(\varphi^+\varphi)^2$ theory.
The field expansion of the second quantized field $\hat\varphi$ reads

\begin{equation}
\hat\varphi(x)=\sum_{\stackrel{\rightharpoonup}{k},\pm}\left\{
\frac{1}{\sqrt{2\omega}}
e^{-i\omega t+i\stackrel{\rightharpoonup}{k}\cdot\stackrel{\rightharpoonup}{x}}
\cdot a(\stackrel{\rightharpoonup}{k},+) +\frac{1}{\sqrt{2\omega}}
e^{i\omega t +i\stackrel{\rightharpoonup}{k}\cdot\stackrel{\rightharpoonup}{x}}
\cdot a(\stackrel{\rightharpoonup}{k},-)\right\}
\end{equation}
and we insert it into the expression for the interaction Hamiltonian 

\begin{equation}
H_{int}=\frac{\lambda}{4}\int
d^3\stackrel{\rightharpoonup}{x}\hat\varphi^+(\stackrel{\rightharpoonup}{x},t)^2\hat\varphi 
(\stackrel{\rightharpoonup}{x},t)^2 \ \ \ .
\end{equation}
With this interaction there is only the s-wave scattering in the
partial wave analysis for two charged scalars.
In the partial wave analysis, the delay or advance of the
scattered particles is given by $\frac{d\delta(\omega)}{d\omega}$ , 
where $\delta(\omega)$ is the energy of one of the outgoing particles.
Thus we may write for very short range interaction 

\begin{equation}
\frac{d\delta(\omega)}{d\omega}=0
\end{equation}
In terms of the spatial wave function in the s-wave channel, the phase 
shift is defined such that the s-wave wave function should be of the
form 
\begin{equation}
\cos\delta(\omega)\cdot j_0(k\cdot r)-
\sin\delta(\omega)\cdot n_0(k\cdot r) \ \ \  .
\label{7.4}
\end{equation}
Here $r$ is the relative distance in the center of mass system and
$\stackrel{\rightharpoonup}{k}$ the relative momentum.
The functions $j_0(k\cdot r)$ and $n_0(k\cdot r)$ denote

\begin{equation}
j_0(k\cdot r) = \frac{\sin k\cdot r}{k\cdot r}
\end{equation}
and 
\begin{equation}
n_0(k\cdot r) = -\frac{\cos k\cdot r}{k\cdot r} \ \ \ .
\end{equation}
It should be noticed that the four momentum in the center of mass
system can be space-like in our naive vacuum and 
thus a proper center of mass system does not exist.
However, we may postpone that case for the moment.

The wave function is expressed in a superposition of eq.(\ref{7.4}) as

\begin{equation}
\varphi(\stackrel{\rightharpoonup}{x}_{rel},\stackrel{\rightharpoonup}{x})=
\int c(\omega) e^{-2i\omega t}\left\{\cos\delta (\omega)j_0(k\cdot r)-
\sin\delta(\omega)n_0(k\cdot r)\right\}d\omega
\end{equation}
where $t$ is the time in the center of mass system.

For very short relative distances we have as $r \rightarrow 0$,

\begin{eqnarray}
n_0(k\cdot r)&\rightarrow&-\frac{1}{kr}+O(k\cdot r) \ \ \ ,\nonumber\\
j_0(k\cdot r)&\rightarrow&  \ 1 \ \ .
\end{eqnarray}
We can extract the coefficients of $n_0$ and $j_0$ as the ones to the 
$\frac{1}{r}$-art and constant part of $\varphi$, which can be
estimated as 

\begin{eqnarray}
{\rm coeff. \  of \ } 
\varphi|_{\frac{1}{r}-part}&=&\lim _{r\rightarrow 0}r\varphi\nonumber\\
&=&\int c(\omega)\sin\delta(\omega)\frac{e^{-2i\omega t}}{k}
d\omega
\end{eqnarray}
and
\begin{eqnarray}
{\rm coeff. \  of \ } 
\varphi|_{const-part}&=&\lim_{r\rightarrow
0}(\varphi-\frac{1}{r}\varphi|{}_{\frac{1}{r}-part})\nonumber\\
&=&\int c(\omega)\cos\delta(\omega)e^{-2i\omega t}d\omega \ \ \ .
\end{eqnarray}
For large $|\omega|$ the masses may be ignored and $\omega$ behaves as 

\begin{equation}
|\omega| \sim k \ \ \ .
\end{equation}
By taking $\delta(\omega)=\delta$ which is constant we find a relation 

\begin{equation}
i
\left({\rm coeff. \ of \ }\dot\varphi|_{\frac{1}{r}-part}\right)=
	2\tan \delta\cdot\left({\rm coeff. \ of \ }
	\varphi|_{const-part}\right)\ \ \ .
\label{7.12}
\end{equation}
We then evaluate how the constant phase shift $\delta$ is related to
the coupling constant $\lambda$.

Before going into this evaluation we must construct a wave function
for two particle state by using the $(\pi,\varphi$)-formalism where
$\pi$ denotes conjugate momentum to $\varphi$.
An $N$-scalar wave function have $2^N$ components and each of those
are marked as $\varphi$ or $\pi$ with respect to each of the $N$
particles in the state.
For example a two particle state wave function may be written as 

\begin{equation}
\left(
\begin{array}{cc}
\varphi_{\varphi\varphi}&
	\hskip-2\arraycolsep(\stackrel{\rightharpoonup}{x_1},
	\stackrel{\rightharpoonup}{x_2}; t)\\
\varphi_{\varphi\pi}&
	\hskip-2\arraycolsep(\stackrel{\rightharpoonup}{x_1},
	\stackrel{\rightharpoonup}{x_2}; t)\\
\varphi_{\pi\varphi}&
	\hskip-2\arraycolsep(\stackrel{\rightharpoonup}{x_1},
	\stackrel{\rightharpoonup}{x_2}; t)\\
\varphi_{\pi\pi}&
	\hskip-2\arraycolsep(\stackrel{\rightharpoonup}{x_1},
	\stackrel{\rightharpoonup}{x_2}; t)
\end{array}
\right) \ \ \ .
\end{equation}
The inner product of these wave functions can be defined by extending
the one for single particle states.
The latter reads

\begin{equation}
\left<\left(
\begin{array}{l}
\varphi_\varphi\\
\varphi_\pi
\end{array}
\right)_1\Bigg|
\left(
\begin{array}{l}
\varphi_\varphi\\
\varphi_\pi
\end{array}
\right)_2\right>=
-i\int(\varphi^*_{\pi 1}\varphi_{\varphi 2}-
\varphi^*_{\varphi1}\varphi_{\pi 2}) \ d^3 \stackrel{\rightharpoonup}{x}
\end{equation}
where the subscript 1 and 2 denote two single particle states.
This expression may be extended to the two particle system as 

\begin{equation}
\left<\left(
\begin{array}{l}\varphi_{\varphi\varphi1}\\
		\varphi_{\varphi\pi1}\\
		\varphi_{\pi\varphi1}\\
		\varphi_{\pi\pi1}
\end{array}\right)\Bigg|\left(
\begin{array}{l}\varphi_{\varphi\varphi2}\\
		\varphi_{\varphi_\pi2}\\
		\varphi_{\pi\varphi2}\\
		\varphi_{\pi\pi2}
\end{array}\right)\right>
=\int(\varphi^*_{\varphi\varphi}
	\varphi^*_{\varphi\pi}
	\varphi^*_{\pi\varphi}
	\varphi^*_{\pi\pi})_1
\left(
\begin{array}{cccc}
0&0&0&-1\\
0&0&1&0\\
0&1&0&0\\
-1&0&0&0
\end{array}
\right)
\left(
\begin{array}{l}\varphi_{\varphi\varphi}\\
		\varphi_{\varphi\pi}\\
		\varphi_{\pi\varphi}\\
		\varphi_{\pi\pi}
\end{array}\right)_2
d^3\stackrel{\rightharpoonup}{x_1} \
d^3\stackrel{\rightharpoonup}{x_2} \ \ \ .
\end{equation}

The second quantized filed
$\hat\varphi(\stackrel{\rightharpoonup}{x})$ 
acts on the two particle state as 

\begin{equation}
\hat\varphi(\stackrel{\rightharpoonup}{x})
\left(
\begin{array}{l}
\varphi_{\varphi\varphi}
	(\stackrel{\rightharpoonup}{x_1}\ , \ 
	 \stackrel{\rightharpoonup}{x_2})\\
\varphi_{\varphi\pi}
	(\stackrel{\rightharpoonup}{x_1}\ , \ 
	 \stackrel{\rightharpoonup}{x_2})\\
\varphi_{\pi\varphi}
	(\stackrel{\rightharpoonup}{x_1}\ , \ 
	 \stackrel{\rightharpoonup}{x_2})\\
\varphi_{\pi\pi}
	(\stackrel{\rightharpoonup}{x_1}\ , \ 
	 \stackrel{\rightharpoonup}{x_2})
\end{array}\right)=\left(
\begin{array}{l}
\varphi_\varphi
	(\stackrel{\rightharpoonup}{x_1})\\
\varphi_\pi
	(\stackrel{\rightharpoonup}{x_1})
\end{array}
\right) 
\delta^3(\stackrel{\rightharpoonup}{x}-
	\stackrel{\rightharpoonup}{x_2})+\left(
\begin{array}{l}
\varphi_\varphi
	(\stackrel{\rightharpoonup}{x_2})\\
\varphi_\pi
	(\stackrel{\rightharpoonup}{x_2})
\end{array}
\right)
\delta^3(\stackrel{\rightharpoonup}{x}-
	\stackrel{\rightharpoonup}{x_1})
\ \ \ .
\end{equation}

Thus $\hat\varphi(\stackrel{\rightharpoonup}{x})$ is just a pure
annihilation operator.
On the other hand, the $\hat\pi$-operator acts as

\begin{equation}
\hat\pi(\stackrel{\rightharpoonup}{x})
\left(
\begin{array}{l}
\varphi_{\varphi\varphi}
	(\stackrel{\rightharpoonup}{x_1}\ , \ 
	 \stackrel{\rightharpoonup}{x_2})\\
\varphi_{\varphi\pi}
	(\stackrel{\rightharpoonup}{x_1}\ , \ 
	 \stackrel{\rightharpoonup}{x_2})\\
\varphi_{\pi\varphi}
	(\stackrel{\rightharpoonup}{x_1}\ , \ 
	 \stackrel{\rightharpoonup}{x_2})\\
\varphi_{\pi\pi}
	(\stackrel{\rightharpoonup}{x_1}\ , \ 
	 \stackrel{\rightharpoonup}{x_2})
\end{array}\right)=\left(
\begin{array}{l}
\varphi_\varphi
	(\stackrel{\rightharpoonup}{x_1})\\
\varphi_\pi
	(\stackrel{\rightharpoonup}{x_1})
\end{array}
\right) 
\delta^3(\stackrel{\rightharpoonup}{x}-
	\stackrel{\rightharpoonup}{x_2})+\left(
\begin{array}{l}
\varphi_\varphi
	(\stackrel{\rightharpoonup}{x_2})\\
\varphi_\pi
	(\stackrel{\rightharpoonup}{x_2})
\end{array}
\right)
\delta^3(\stackrel{\rightharpoonup}{x}-
	\stackrel{\rightharpoonup}{x_1})
\ \ \ .
\end{equation}

Next we consider the Hermitean conjugate operator 
$\hat\varphi(\stackrel{\rightharpoonup}{x})^+$
which should act on the states in such away that it has the same 
effect as $\varphi(\stackrel{\rightharpoonup}{x})$ in the ket.
Thus

\begin{equation}
\hat\varphi^+(\stackrel{\rightharpoonup}{x})
\left(
\begin{array}{l}
	\varphi_\varphi(\stackrel{\rightharpoonup}{x_1})\\
	\varphi_\pi(\stackrel{\rightharpoonup}{x_1})
\end{array}\right)=\left(
\begin{array}{lcl}
\varphi_{\varphi\varphi}
	(\stackrel{\rightharpoonup}{x_1}\ , \ 
	 \stackrel{\rightharpoonup}{x_2})&=&0\\
\varphi_{\varphi\pi}
	(\stackrel{\rightharpoonup}{x_1}\ , \ 
	 \stackrel{\rightharpoonup}{x_2})&=&
	\varphi_\varphi(\stackrel{\rightharpoonup}{x_1})\delta^3
	(\stackrel{\rightharpoonup}{x_2}-\stackrel{\rightharpoonup}{x})\\
\varphi_{\pi\varphi}
	(\stackrel{\rightharpoonup}{x_1}\ , \ 
	 \stackrel{\rightharpoonup}{x_2})&=&0\\
\varphi_{\pi\pi}
	(\stackrel{\rightharpoonup}{x_1}\ , \ 
	 \stackrel{\rightharpoonup}{x_2})&=&
	\varphi_\pi(\stackrel{\rightharpoonup}{x_1})\delta^3
	(\stackrel{\rightharpoonup}{x_2}-\stackrel{\rightharpoonup}{x})
\end{array}\right)
\end{equation}

Therefore starting from zero particle state $\varphi_0$ we obtain

\begin{eqnarray}
\left(\hat\varphi^+
	\left(\stackrel{\rightharpoonup}{x}\right)\right)^2\varphi_0&=&
	\hat\varphi(\stackrel{\rightharpoonup}{x})^+
	\left(
\begin{array}{l}
	\varphi_\varphi(\stackrel{\rightharpoonup}{x_1})=0\\
	\varphi_\pi(\stackrel{\rightharpoonup}{x_1})=
		\delta^3(\stackrel{\rightharpoonup}{x}-x_1)
\end{array}\right)\nonumber\\
&=&\left(
\begin{array}{l}
	\varphi_{\varphi\varphi}
	(\stackrel{\rightharpoonup}{x_1} \ , \
 	\stackrel{\rightharpoonup}{x_2})=0\\
	\varphi_{\varphi\pi}
(\stackrel{\rightharpoonup}{x_1} \ , \
 	\stackrel{\rightharpoonup}{x_2})=0\\
	\varphi_{\pi\varphi}
	(\stackrel{\rightharpoonup}{x_1} \ , \
 	\stackrel{\rightharpoonup}{x_2})=0\\
	\varphi_{\pi\pi}
	(\stackrel{\rightharpoonup}{x_1} \ , \
 	\stackrel{\rightharpoonup}{x_2})=
	\delta^3(\stackrel{\rightharpoonup}{x}-
	\stackrel{\rightharpoonup}{x_1})
	\delta^3(\stackrel{\rightharpoonup}{x}-
	\stackrel{\rightharpoonup}{x_2})
\end{array}\right)
\end{eqnarray}

It is now clear how the Hamiltonian operator $\hat H$ should be 
expressed by the second quantized fields 
$\hat\varphi$ and $\hat\varphi^+$ .
The free Hamiltonian operator 

\begin{equation}
\hat H_{free}=\int\left\{
	\pi^+(\stackrel{\rightharpoonup}{x})
	\pi(x)+
	(\bigtriangledown\varphi^+(\stackrel{\rightharpoonup}{x}))\cdot
	(\bigtriangledown\varphi(\stackrel{\rightharpoonup}{x}))\right\}
	d^3\stackrel{\rightharpoonup}{x}
\end{equation}

acts on the two particle states as 

\begin{equation}
\hat H_{free}\left(
\begin{array}{l}
\varphi_{\varphi\varphi}\\
\varphi_{\varphi\pi}\\
\varphi_{\pi\varphi}\\
\varphi_{\pi\pi}
\end{array}\right)=\left(
\begin{array}{llll}
0&i&i&0\\
-i\Delta&0&0&i\\
-i\Delta&0&0&i\\
0&-i\Delta&-i\Delta&0
\end{array}\right)\left(
\begin{array}{l}
\varphi_{\varphi\varphi}\\
\varphi_{\varphi\pi}\\
\varphi_{\pi\varphi}\\
\varphi_{\pi\pi}
\end{array}\right) \ \ \ .
\end{equation}

For the interaction Hamiltonian 

\begin{equation}
\hat H_{int}=\frac{\lambda}{8}\int
(\varphi(\stackrel{\rightharpoonup}{x})^+)^2
\varphi(\stackrel{\rightharpoonup}{x})^2 d^3\stackrel{\rightharpoonup}{x}
\end{equation}

we have 

\begin{equation}
\hat H_{int}\left(
\begin{array}{l}
\varphi_{\varphi\varphi}
	(\stackrel{\rightharpoonup}{x_1},
		\stackrel{\rightharpoonup}{x_2} ; t)\\
\varphi_{\varphi\pi}
	(\stackrel{\rightharpoonup}{x_1},
		\stackrel{\rightharpoonup}{x_2} ; t)\\
\varphi_{\pi\varphi}
	(\stackrel{\rightharpoonup}{x_1},
		\stackrel{\rightharpoonup}{x_2} ; t)\\
\varphi_{\pi\pi}
	(\stackrel{\rightharpoonup}{x_1},
		\stackrel{\rightharpoonup}{x_2} ; t)
\end{array}\right)=\frac{\lambda}{8}\left(
\begin{array}{l}
0\\
0\\
0\\
\delta^3(\stackrel{\rightharpoonup}{x_1}-\stackrel{\rightharpoonup}{x_2})
	\varphi_{\varphi\varphi}
	(\stackrel{\rightharpoonup}{x_1},
		\stackrel{\rightharpoonup}{x_2} ; t)
\end{array}\right) \ \ \ .
\end{equation}

In an energy eigenstate the average of $\hat H_{int}$ is given as 

\begin{eqnarray}
\left<\hat H_{int}\right>&=&\left<\left(
\begin{array}{l}
\varphi_{\varphi\varphi}\\
\varphi_{\pi\pi}\\
\varphi_{\pi\varphi}\\
\varphi_{\pi\pi}
\end{array}\right)\Bigg|\hat H_{int}\Bigg|\left(
\begin{array}{l}
\varphi_{\varphi\varphi}\\
\varphi_{\varphi\pi}\\
\varphi_{\pi\varphi}\\
\varphi_{\pi\pi}
\end{array}\right)\right> \Bigg/ \left<\left(
\begin{array}{l}
\varphi_{\varphi\varphi}\\
\varphi_{\varphi\pi}\\
\varphi_{\pi\varphi}\\
\varphi_{\pi\pi}
\end{array}\right)\Bigg|\left(
\begin{array}{l}
\varphi_{\varphi\varphi}\\
\varphi_{\varphi\pi}\\
\varphi_{\pi\varphi}\\
\varphi_{\pi\pi}
\end{array}\right)\right>\nonumber\\
&=&<\frac{\lambda}{8\omega^2}\delta^3
(\stackrel{\rightharpoonup}{x_1}-\stackrel{\rightharpoonup}{x_2})>
\end{eqnarray}

where $<>$ denotes to take the expectation value by the eigenfunction
of $\hat H_{free}$ with the eigenvalue $\omega$

In order to estimate the phase shift caused by the interaction 
$\frac{\lambda}{8}(\varphi^+\varphi)^2$
we consider an incoming wave packet with zero angular momentum
corresponding to the $s$-wave, which has a distance $L$ in the radial
direction .
It is expressed by 
$
ch_0(k\cdot r) $
as 
$ h_0(k\cdot r)\sim\frac{e^{-ik\cdot r}}{k\cdot r}$
where a coefficient $c$ is determined by 

\begin{equation}
\int^{R+\frac{L}{2}}_{R-\frac{L}{2}} |
ch_0(k\cdot r)|^2 \ \cdot \ 4\pi r^2dr=\frac{1}{(2\omega)^2} \ \ \ .
\end{equation}

Here $R$ denotes the radial position of the center
and we use normalization so that after dropping center of mass motion
it is given by 
$-i\int(\pi^*\varphi-\varphi^*\pi)d^3\stackrel{\rightharpoonup}{x}=1$.
For $\omega=k$ it gives

\begin{equation}
c=\frac{1}{4\sqrt{\pi L}} \ \ \ .
\end{equation}
While the wave passes with time $L$ the square of the wave function 
$|\varphi_{\varphi\varphi}|^2$ is given by 

\begin{equation}
|cj_0|^2=|c|^2=\frac{1}{16\pi L} \ \ \ .
\end{equation}
Thus the expectation value of $\hat H_{int}$
in the period $L$ is evaluated as 

\begin{equation}
\frac{\lambda}{8}
\int\varphi^*_{\varphi\varphi}\varphi_{\varphi\varphi}
d^3\stackrel{\rightharpoonup}{x}=
\frac{\lambda}{8}|c|^2=
\frac{\lambda}{128\pi L}
\end{equation}
and so the phase shift of the wave function on the average is given by 

\begin{equation}
\delta=-\frac{\lambda}{128\pi L}L=-\frac{\lambda}{128\pi} \ \ \ .
\end{equation}
Using this relation to the boundary condition eq.(\ref{7.12}), 
it is expressed in terms of $\lambda$ 

\begin{equation}
{\rm coeff. \ of \ }
\dot\varphi_{\varphi\varphi}|_{\frac{1}{r}-part}=-
\frac{\lambda}{64\pi}
({\rm coeff. \ of \ }\varphi_{\varphi\varphi}|_{const-part})
\label{7.30}
\end{equation}
where small $\delta$ approximation is used.
Furthermore if one use the relation in the center of mass frame 

\begin{equation}
\dot\varphi_{\varphi\varphi}=\varphi_{\varphi\pi}=\varphi_{\pi\varphi}
\end{equation}
we can write the boundary condition  eq.(\ref{7.30}) as

\begin{equation}
{\rm coeff. \ of \ }
\varphi_{\varphi\pi}|_{\frac{1}{r}-part}=-
\frac{\lambda}{64\pi}\cdot
({\rm coeff. \ of \ }\varphi_{\varphi\varphi}|_{const-part}) \ \ \ .
\end{equation}

In general one may rewrite local interactions as boundary conditions
in the similar manner to the above example.

\section{The analyticity and sheet structure}

We argue in this section that it is natural to think of analytic wave
functions in the naive vacuum world.
The only exception should be branch-point singularities 
where one of the distances 
$r_{ik}=
	\sqrt{(\stackrel{\rightharpoonup}{x_i}-
		\stackrel{\rightharpoonup}{x_k})^2}$ 
between a couple of particle vanishes.

However, as will be shown below the scattering of negative energy
particles on positive energy ones leads to bad singularities in the
wave functions.
To argue for these singularities we start from scattering of two particles 
with ordinary time-like total four momentum $p^\mu_1+p^\mu_2$ .
The local interaction, typically in $s$-waves, leads to a scattering
amplitude expressed by $\eta_0(kr)=-\frac{\cos kr}{kr}$ which 
is singular at $k\cdot r=0$.
Here $k$ and $r$ denote the relative momentum and distance.
It has a pole in $k\cdot r$ but has a square root branch point as a
function of 
$(kr)^2\propto(p_1+p_2)^2(x_1-x_2)^2-[(p_1+p_2)(x_1-x_2)]^2$ .
Then we analytically continue $p^\mu_1+p^\mu_2$ into space-like one.
This is what one can achieve in the naive vacuum world.
When the two formal times $x^0_1$ and $x^0_2$ are equal $x^0_1=x^0_2$
we have the true wave function.
Even under this restriction $x^0_1=x^0_2$ , in an appropriate
reference frame 
we can uphold the zero of $k\cdot r$ and thus branch point
singularities for some real $x^\mu_1-x^\mu_2$ and $p^\mu_1+p^\mu_2$.

One would be worried that such singularities might challenge
causality.
However this is not the case because of the following reason:
The singularity only appears in an infinitely sharp eigenstate of
$p^\mu_1+p^\mu_2$. 
The set up of such a state in principle require so extended apparatus
that one can no longer test causality.
In fact, suppose that we set $p^\mu_1+p^\mu_2$ with a finite spread
$\Delta(\stackrel{\rightharpoonup}{p_1}+\stackrel{\rightharpoonup}{p_2})$
then apparatus have to be no more extended than 
$\frac{\hbar}{\Delta(\stackrel{\rightharpoonup}{p_1}+
		\stackrel{\rightharpoonup}{p_2})}$ 
in $\stackrel{\rightharpoonup}{x}$-space.
But in this case the singularity also gets washed out so as to go into 
the complex plane.

However, even if this is the case, it is unavoidable that the
distances
$r_{ik}=
	\sqrt{(\stackrel{\rightharpoonup}{x_i}-
		\stackrel{\rightharpoonup}{x_k})^2}$
can go to zero even if 
$\stackrel{\rightharpoonup}{x_i}-\stackrel{\rightharpoonup}{x_k}$
does not go to zero.
Thus there will be branch point singularities.

Next we argue how many sheets there are in the $N$ particle scattering.
The wave functions are prepared and thus whether it is analytic or not
may depend on the preparation.
However even if we prepare an analytic wave function there may be
unavoidable branch-point singularities when the distance $r_{ik}$ goes 
to zero for nonzero complex distance vector.
These unavoidable singularities come in from the interactions because 
$\eta_0$ is singular at $r=0$.

If we consider the scattering of two particles, it is described by the 
phase shift.
The wave functions of the scattered particles are described in terms
of 
$\jmath_\ell(kr_{12}){\cal Y}^\ell_m
(\stackrel{\rightharpoonup}{x_1}-\stackrel{\rightharpoonup}{x_2})$
and 
$\eta_\ell(kr_{12}){\cal Y}^\ell_m
(\stackrel{\rightharpoonup}{x_1}-\stackrel{\rightharpoonup}{x_2})$,
where $r_{12}$ is the distance of the two particles 1 and 2 and 
$\stackrel{\rightharpoonup}{x_1}$ , 
$\stackrel{\rightharpoonup}{x_2}$ their positions.
The spherical harmonics ${\cal Y}^\ell_m$ are defined as

\begin{equation}
{\cal Y}^\ell_m(\stackrel{\rightharpoonup}{x_1}-x_2)=r^\ell y^\ell_m
\end{equation}
where $y^\ell_m$ are the usual $r_{12}$-independent spherical
harmonics. 
When there is no scattering, it is expressed by 
$\jmath_\ell{\cal Y}^\ell_m$
and there will be no branch point singularities. 
However if there is scattering, the functions

\begin{equation}
\eta_\ell(k\cdot r_{12}){\cal Y}^\ell_m
\end{equation}
come into the wave function which are odd in $r_{12}$ so that the
branch point singularities appear.
Here $r_{12}$ is a simple square root and thus it has two sheets on 
which the wave functions differ by the sign of 
$\eta_\ell(k\cdot r_{12}){\cal Y}^\ell_m$ .
These parts of the wave functions are given by the interaction as was
discussed in the preceding section.
In particular the switch of the sign can be achieved by keeping the
$\jmath_\ell{\cal Y}^\ell_m$ terms fixed but changing the sign  of the 
interaction, $\lambda\rightarrow -\lambda$.

If we consider $N$-particle system, there are $\frac{1}{2}N(N+1)$ pairs 
of particles of which distance is given by 
$r_{ik}=
	\sqrt{(\stackrel{\rightharpoonup}{x_i}-
		\stackrel{\rightharpoonup}{x_k})^2}$.
Since we consider just 2 sheets with respect to each of these
$\frac{1}{2}N(N+1)$ pairs, there appear $2^{\frac{1}{2}N+1)}$ sheets
altogether.

We consider the case that the incoming waves scatter in all possible
combinations with each other.
Thus a wave scattered $\eta_{ik}$ times by particles $i$ and $k$
changes sign 

\begin{equation}
\sum_{(i,k)}\eta_{ik}\cdot S_{ik} \ \ \ (i\neq k)
\end{equation}
times when going from the original sheet to the one characterized by 
$\{S_{ik}\}$.
Here $S_{ik}$ is understood to be $S_{ik}=0$ for the first sheet with
respect to $r_{ik}$ and $S_{ik}=1$ for the second sheet.
So the piece of the wave function changes as 

\begin{equation}
\psi_{piece} \rightarrow
(-1)^{\sum_{(i,k)}\eta_{ik}S_{ik}}\psi_{piece} \ \ \ .
\end{equation}
Obviously this sign changes separate the $2^{\frac{1}{2}N(N+1)}$
sheets and this is the minimal number of sheets which gives the wave
function with minimal singularities of the branch point.

Typically even if the incoming waves have other singularities, we can
continue to other sheet with respect to the $r_{ik}$ distance as
will be discussed in next section.

\section{The CPT-like theorem}

In this section we first consider how the usual CPT symmetry operation 
acts on the naive vacuum world.
As will be explained below the naive vacuum is brought into the
totally different state if the usual CPT is applied and thus it is
spontaneous by broken.
We shall then propose a new CPT-like symmetry which leaves the naive
vacuum invariant.

In order to apply the usual CPT operation, we describe the naive
vacuum from the true vacuum viewpoint.
In the naive vacuum viewpoint there are no particles and/or no holes
in all the single particle states with positive energies as well as
negative ones.
Even if we consider them from the true vacuum point of view there are
no particles in the positive energy states.
However the negative energy states are described as the ones with
minus one hole for bosons and one hole for fermions.
If we use the states with only the positive energies there are minus one
and one anti-particles for bosons and fermions respectively.

The usual CPT operation acts on the naive vacuum described by the true 
vacuum viewpoint in the following way : it brings into the positive
energy states with minus one  boson and one fermions.
Theses states should be translated back to the ones in terms of the
naive vacuum viewpoint.
All the states with positive energies as sell as negative ones are
filled by minus one  boson and one fermions.
These are completely new states relative to the naive vacuum and we
may say that the usual CPT symmetry is spontaneously broken.

We now ask the question if we could invent a CPT-like
operation for which the naive vacuum does not represent such a
spontaneous breakdown.

We shall indeed answer this question affirmatively :
we propose a new strong reflection in which the inversion of the
operator order contained in the usual strong reflection is excluded.
We then compose it with an operation of analytic continuation which
will be presented below.

At first one might think that a primitive strong reflection on a state
in the naive vacuum world be a good symmetry.
This reflection is defined so as to shift the sign of any Lorentz
tensor by $(-1)^{\rm rank}$ .
Thus the Hamiltonian gets a sign shift $H\rightarrow -H$ under the
primitive strong reflection.
Indeed this expectation works for the free theory.
However, looking at the interaction described by the boundary
condition in eq.(\ref{7.12}), we see that under the time reversal
$t\rightarrow -t$ which is contained in the strong reflection the left 
hand side changes the sign while the right hand side does not.
This means that action of the naive strong reflection on a state in the 
Fock space in the naive vacuum world which obeys the condition 
eq.(\ref{7.12})brings into a state which does not obey it.
If we should make the symmetry a good one we might do it so by
changing $\tan\delta\rightarrow -\tan\delta$.
It would be achieved by changing sign on the coupling constant
$\lambda$ according to the formula eq.(\ref{7.30}).
However the coupling constant $\lambda$ is one of the fundamental
quantities and it cannot change sign.
Thus the naive strong reflecting without operator order inversion is not 
a good symmetry for the interaction $H_{int}$.

We may look for a modification of the Fock space state of the model
that could provide a sign shift of the interaction.
By combining the modification with the simple strong reflection we
could get a symmetry of the laws of nature in the naive vacuum picture.

For finding this modification it would provide an important hint that
in the relation eq.(\ref{7.12}) there is a factor $\frac{1}{r_{ik}}$ and that
the distance $r_{ik}$ is a square root
$r_{ik}=
\sqrt{(\stackrel{\rightharpoonup}{x_i}-
	\stackrel{\rightharpoonup}{x_k})^2}$.
This distance has a branch point at $r_{ik}=0$ and $\infty$ as an
analytic complex function.
In the analytic continuation it has two sheets on which $r_{ik}$ takes 
on the opposite values

\begin{equation}
r_{ik}|_{\rm one \ sheet} = -r_{ik}|_{\rm the \ other \ sheet} \ \ \ .
\end{equation}

This sign ambiguity is just what we need, provided we postulate that 
our new CPT-like operation should contain the analytic continuation
onto the opposite sheet in addition to the simple strong reflection.
By the former change we get $r_{ik}\rightarrow -r_{ik}$ and it makes
the boundary condition eq.(\ref{7.12}) turn out invariant under the full
operation.
Thus the new CPT-like operation consists in 
A) simple strong reflection composed with B) $r_{ik}\rightarrow
-r_{ik}$ provided by analytic continuation.

In order to get the sign shift $r_{ik}\rightarrow-r_{ik}$ for the
interactions between all possible pairs of particles in the naive
vacuum world it is needed to make the analytic continuation so that it
goes to the opposite sheet with respect to every pair of particles
$i,k$.
In fact we restrict our attention to such a wave function that it has
only the branch points due to the square roots in $r_{ik}$.
With such an assumption for the wave functions we may define a
procedure B) by which a wave function $\psi$ is replaced by a new one
for the same values of 
$(\stackrel{\rightharpoonup}{x_1},
\stackrel{\rightharpoonup}{x_2},\ldots,
\stackrel{\rightharpoonup}{x_N}))$
but on the different sheet.

All this means that we propose a replacement for the usual CPT-theorem 
in order to avoid spontaneous breakdown by the naive vacuum.

We consider the CPT-like theorem
by taking the simple case of $(\varphi^+\varphi)^2$ theory.
It should be valid under very broad conditions.

In the $(\varphi^+\varphi)^2$ theory in the naive vacuum world there
is a symmetry under the following combined operation of A) and B) :

A) Strong reflection without operator inversion.\\
We make a replacement in the wave function of all position variables 
$\stackrel{\rightharpoonup}{x_1}x_2,\ldots,x_N$ by their opposite
ones.

\begin{equation}
\psi_{\pi\varphi\ldots}
	(\stackrel{\rightharpoonup}{x_1},\ldots,
	 \stackrel{\rightharpoonup}{x_N})\rightarrow
	(-1)^{\#\pi-{\rm indices}}
\psi_{\pi\varphi\ldots}
	(-\stackrel{\rightharpoonup}{x_1},\ldots,
	 -\stackrel{\rightharpoonup}{x_N})
\end{equation}
whereby replacements $\stackrel{\rightharpoonup}{p_i}\rightarrow
-\stackrel{\rightharpoonup}{p_i}$ are performed.
In this operation energy and thereby time shift sign.
This means that $\psi_{\varphi\ldots\pi\ldots\varphi}$ components have 
a minus sign for each $\pi$

\begin{equation}
\psi_{\varphi\varphi\pi\varphi\ldots\pi_N}
	(\stackrel{\rightharpoonup}{x_1},\ldots,
	 \stackrel{\rightharpoonup}{x_N})\rightarrow 
	(-1)^{\#(\pi-{\rm indices})}
\psi_{\varphi\varphi\pi\varphi\ldots\pi_N}
	(-\stackrel{\rightharpoonup}{x_1},\ldots,
	  \stackrel{\rightharpoonup}{x_N}) \ \ \ . 
\end{equation}
this transformation A) may be described as strong reflection (S.R.)
though without including the operator inversion rule.
It should be noticed that this operation A) does not keep the inner
product invariant.

B) Analytic continuation of the wave function onto the other sheet
on which all the distances $r_{ik}$ shift sign.

The wave function 
$\psi_{\varphi\pi\ldots}
(\stackrel{\rightharpoonup}{x_1},\ldots,\stackrel{\rightharpoonup}{x_N})$ 
is analytically continued onto another sheet over the complex 3N
dimensional configuration space but with the same values of the
arguments
$\stackrel{\rightharpoonup}{x_1},\ldots,\stackrel{\rightharpoonup}{x_N}$ 
.
The continued sheet is the one on which the relative distances 
$r_{ik}=\sqrt{(\stackrel{\rightharpoonup}{x}_i-
	\stackrel{\rightharpoonup}{x}_k)^2}$
are shifted their sign relative to the first sheet.

This CPT-like theorem presupposes to choose boundary conditions for long
distances in the following way :

\begin{enumerate}
\renewcommand{\labelenumi}{\arabic{enumi})}
\item We suppose that in the scattering at $t\rightarrow -\infty$ we have 
ingoing waves while at $t\rightarrow\infty$ outgoing waves. 
Here ingoing and outgoing waves are interpreted according to the
directions of velocities but not momenta.
\item For the negative energy pairs we choose, instead of convergent
boundary conditions anticonvergent boundary conditions, which means
that total energy is negative.
Some explanation is in order.
\end{enumerate}

First of all we call attention to the relations between the velocity
$\stackrel{\rightharpoonup}{\upsilon}$ and momentum
$\stackrel{\rightharpoonup}{p}$.
For positive energy particles $\stackrel{\rightharpoonup}{\upsilon}$
and $\stackrel{\rightharpoonup}{p}$ point in the same direction while
for the negative energy particles it is opposite.
Whether the wave is incoming or outgoing is defined by the velocity.

The operation A) brings an extended bound state over a finite region
into a configuration with a finite extension.
If a bound state wave function behaves as $e^{-kr}$, it may blows up
as $e^{kr}$ after the analytic continuation B).
This is an obvious consequence of changing the sign of all the formal
expression 
$r_{ik}=\sqrt{(\stackrel{\rightharpoonup}{x}_i-
		\stackrel{\rightharpoonup}{x}_k)^2}$
for the relative distances.
Then in order to be able to define the condition on the blow up state, 
we have to analytically continue back these distances to the sheet
with no sign shift relative to the original sign.
Thus we can summarize in the following way :
For positive energy we use ordinary bound state condition while for
negative energy the anti-bound condition is required.

It should be remarked that both energy and momentum change the sign,
but not the velocity.

We may list the properties of operations A) and B) : 

Properties of A)

1) transformations of variables

\begin{eqnarray}
&&\stackrel{\rightharpoonup}{x}
\rightarrow -\stackrel{\rightharpoonup}{x} \ \ \ , \ \ \ 
t\rightarrow -t\nonumber\\
&&\stackrel{\rightharpoonup}{p}
\rightarrow-\stackrel{\rightharpoonup}{p} \ \ \ , \ \ \ 
E\rightarrow -E \nonumber\\
&&\stackrel{\rightharpoonup}{\upsilon}\rightarrow 
\stackrel{\rightharpoonup}{\upsilon}\nonumber\\
&&i\rightarrow  i ({\rm  no \  complex \ conjugation})\nonumber
\end{eqnarray}

2) boundary conditions\\
$s$-matrix type boundary condition, that is, ingoing at $t=-\infty$
and outgoing at $t=\infty$ are kept fixed.
In terms of velocity,

\begin{eqnarray}
{\rm ingoing \ wave} &\rightarrow& {\rm outgoing \ wave}\nonumber\\
(\stackrel{\rightharpoonup}{x}\cdot
	\stackrel{\rightharpoonup}{\upsilon}<0)&&
(\stackrel{\rightharpoonup}{x}\cdot
	\stackrel{\rightharpoonup}{\upsilon}>0)\nonumber
\end{eqnarray}

3) particle and hole are not exchanged

\begin{equation}
{\rm particle}\rightarrow{\rm particle} \ , \ 
{\rm hole}\rightarrow{\rm hole} \ \ \ .
\nonumber
\end{equation}

Properties of B)

The operation of the analytic continuation in B) is performed on the
maximally analytic wave function
$\psi(\stackrel{\rightharpoonup}{x_1},
\stackrel{\rightharpoonup}{x_2}\ldots,\stackrel{\rightharpoonup}{x_N})$ 
for any fixed number $N$ of particles.
In this way the number of particles $N$ is conserved.

A continuous function 
$\psi(\stackrel{\rightharpoonup}{x_1},
\stackrel{\rightharpoonup}{x_2}\ldots,\stackrel{\rightharpoonup}{x_N})$ 
of variables $(\stackrel{\rightharpoonup}{x_1},
\stackrel{\rightharpoonup}{x_2}\ldots,\stackrel{\rightharpoonup}{x_N})$
conceived of as complex variables is analytically continued along a
path which avoids the branch points.
It is lead onto a sheet characterized by the change of the sign for
all the $r_{ik}$ relative to the starting one.

In a typical scattering the scattered wave function will
change sign under the operation.
In the expression in terms of the partial wave of the wave function 

\begin{eqnarray}
&&\varphi_{\varphi\varphi}\propto
	\cos\delta\jmath_0(k\cdot r)-\sin\delta n_0(k\cdot r)  \ \ \ , 
\nonumber\\
&&\varphi_{\varphi\pi}\propto
	\omega\cos\delta\jmath_0(k\cdot r)-
	\omega\sin\delta n_0(k\cdot r)
\end{eqnarray}
we make analytic continuation $r\rightarrow -r$ and 
$n_0 \rightarrow -n_0$ while $\jmath_0\rightarrow \jmath_0$ so that
the scattered wave given by $n_0$ changes sign.

\section{Proof of the CPT-like theorem}

To prove the CPT-like theorem discussed in last section we first
remark that it is true for free theory : 
In this case the operation A) may be brought upon in momentum space and 
both energy and momentum are inverted for every particles.
Since free wave function is analytic and it is the same on all the
sheets.
Thus the continuation operation B) has no effect in the free case.

If there are interactions the strong reflection A) does not keep
invariant the boundary conditions at the meeting point

\begin{equation}
i({\rm coeff. \ of \ }
\varphi_{\varphi\varphi\pi\ldots\varphi}|_{\frac{1}{r}-{\rm part}})=
	2\tan\delta\cdot({\rm coeff. \ of \ }
	\varphi_{\varphi\varphi\pi\ldots\varphi}|_{\rm const.part}) 
	\ \ \ .
\label{10.1}
\end{equation}
with $\tan\delta=-\frac{\lambda}{64\pi}$
because the left hand side changes the sign under $t\rightarrow -t$.
Thus after the operation A) the condition is satisfied with opposite
sign $\tan\delta\rightarrow-\tan\delta$.
On the other hand with analytic continuation under B) the equation of
motion in the $r_{ik}\neq 0$ regions remain satisfied with the
eigenvalue $-E$.
However by the operation A) the solution with $E$ becomes the one with 
$-E$.
But on return the $r_{ik}$'s have changed sign.
Thus after both A) and B) operation not only the equation of motion is 
fulfilled but also the boundary conditions eq.(\ref{10.1}) are satisfied with
the original correct sign.

Let us remark on the boundary condition at large
$\stackrel{\rightharpoonup}{x_i}$'s.
In this case as long as the scattering is negligible the $\jmath_\ell$ 
expansions works which is even in $r=r_{ik}$
in the sense that if one wants to use spherical harmonics you need

\begin{equation}
{\cal Y}^m_\ell
(\stackrel{\rightharpoonup}{x_i}-\stackrel{\rightharpoonup}{x_\ell})=r^\ell 
y^m_\ell\left(\frac{\stackrel{\rightharpoonup}{x_i}-
	\stackrel{\rightharpoonup}{x_\ell}}{r}\right) \ \ \ .
\end{equation}
The final expansion turns out even in $r$ when no $n'_\ell$s are
present.
Thus the wave function is the same on all sheets and the
operation B) is unaffected.
When interaction is switched on in an incoming wave, the
sign is changed for a piece of scattered wave scattered an odd 
number of times. 
However, the form of the waves, either 
$e^{ik|\stackrel{\rightharpoonup}{x}|}$ or 
$e^{-ik|\stackrel{\rightharpoonup}{x}|}$, is not changed by the analytic
continuation  B).
Thus the velocity-wise boundary condition can be settled as if there
were only the operation A) in the CPT-like theorem.
Under A) we have the shifts 
$\stackrel{\rightharpoonup}{x}\rightarrow
-\stackrel{\rightharpoonup}{x} \ , \ 
\stackrel{\rightharpoonup}{p}\rightarrow
-\stackrel{\rightharpoonup}{p} \ , \ 
E\rightarrow -E$ and thus 
$\stackrel{\rightharpoonup}{\upsilon}=
	\frac{\stackrel{\rightharpoonup}{p}}{E}$
is not.
Therefore 
$\stackrel{\rightharpoonup}{x}\cdot
\stackrel{\rightharpoonup}{\upsilon}\rightarrow-
\stackrel{\rightharpoonup}{x}\cdot
\stackrel{\rightharpoonup}{\upsilon}$.

In the $S$-matrix conditions the ingoing wave packets at $t=-\infty$
have
$\stackrel{\rightharpoonup}{x}\cdot\stackrel{\rightharpoonup}{\upsilon}<0$ 
while the outgoing ones have 
$\stackrel{\rightharpoonup}{x}\cdot\stackrel{\rightharpoonup}{\upsilon}>0$.
Combining with $t\rightarrow -t$, this condition remains the same
under A) and thus under the whole CPT-like operation in the naive
vacuum world.

To suggest that this CPT-like theorem is very general we sketch the
method how to construct a general proof.
First operation A) is in fact strong reflection but without the
operation of order change of the creation and annihilation operators.
It is strongly suggested to perform an analytic continuation of a
Lorentz-boost to an imaginary rapidity

\begin{equation}
\eta=i\pi
\end{equation}
where the boost velocity is given by 
$\upsilon_{boost}=\tanh \eta$.
It is followed by a $\pi$ rotation around the boost axis.

It may be written, for simplicity, in terms of the matrix elements of
the Lorentz transform as 

\begin{equation}
\Lambda^\mu\nu=\left(
\begin{array}{cccc}
\cosh\eta&\sinh\eta&0&0\\
-\sinh\eta&\cosh\eta&0&0\\
0&0&1&0\\
0&0&0&1\\
\end{array}\right) \ \ \ .
\end{equation}
However it is not as simple as this.
In order to perform a Lorentz-boost of a wave function we have to
live with the problem that simultaneity is not an absolute concept.
To obtain the wave function in another reference frame, it is required 
that one time develops the state of different positions in a different 
way.
The best way may be to use a formalism in which different times are
allowed for each particles.
Such a formalism is possible locally since the interactions are so
local that the theory is free and we can time develop one particle but 
not others.

In the neighborhood of
$(\stackrel{\rightharpoonup}{x_1},\ldots,\stackrel{\rightharpoonup}{x_N})$ 
for $N$ particles with
$\stackrel{\rightharpoonup}{x_i}\neq\stackrel{\rightharpoonup}{x_\jmath}$ 
we may treat them as free particles for a certain finite time.
Then up to the final time we consider that each particles have
individual times $t_1,\ldots,t_N$.
It may be allowed to boost them with small $\eta$.
For boosts with large $\eta$ the Lorentz transformations will involve
coinciding $\stackrel{\rightharpoonup}{x_k}$'s so that the interaction 
caused by singularities will be associated.

We would like to suggest that under a properly performed Lorentz-boost 
with the imaginary rapidity $\eta=i\pi$ we may be able to let all the
distances 
$r_{ik}=
\sqrt{(\stackrel{\rightharpoonup}{x_i}-
\stackrel{\rightharpoonup}{x_k})^2}$
go onto the second sheet.
This means that the analytic continuation of Lorentz boosts and $\pi$
rotation leads not only to the A operation but also to B).
Thus our CPT-like theorem turns out to be the analytic continuation of 
the Lorentz transformations.

To see typical sheet structure we consider the two particle case in which
they move freely in the classical approximation and pass each other at 
some finite distance or meet at a point.
For simplicity we take a situation in which one of the particles
passes through the origo of the space time coordinate system.
When making a pure Lorentz transformation this particle will continue
to cut the $t'=0$ simultaneity hyperplane in the same event.
This means that the distance between the passage events on the
simultaneity hyperplane of the two particles is given for all
reference frames by the one from the origo to the passage event of
number two particle.

For simplicity we shall also choose the initial coordinate system in
such a way that the particle number two is at rest.
Then the distance between the two particles at zero time $t'=0$ in the 
boosted coordinate system is equal to the Lorentz contracted length
of the stick of which one end is at the origo in the original frame.
Let us suppose that the stick relative to the $x$-axis in the original 
frame have the longitudinal length component $\ell_{''}$ and the
transverse one $\ell_\bot$.
The coordinates of the three space in the original coordinate system
of the particle 2 passage reads

\begin{equation}
\stackrel{\rightharpoonup}{x}=
	(\ell_{''} \ , \ \ell_\bot \ , \ 0)
\end{equation}
by choosing that the $x-y$ plane contains the stick.
The original length from the point $\stackrel{\rightharpoonup}{x}=0$
to $\stackrel{\rightharpoonup}{x}=(\ell_{''} \ , \ \ell_\bot \ , \ 0)$ 
is given by $\ell=\sqrt{\ell^2_{''}+\ell^2_\bot}$ and gets Lorentz
contracted to be 

\begin{equation}
\ell_{(\rm new \ frame)}=\sqrt{\ell^2_\bot+\ell^2_{''}(1-\upsilon^2)}
\end{equation}
where $\upsilon$ is the boost velocity along the $x$-axis in the new
frame. 
In terms of rapidity it reads

\begin{equation}
\ell_{(\rm new \ frame)}=\sqrt{\ell^2_\bot+\ell^2_{''}(1-\tanh^2\eta) }
\end{equation}
This can be written as the factorized radicant form

\begin{equation}
\ell_{(\rm new \ frame)}=\sqrt{(\ell^2_\bot -i\frac{\ell}{\cosh\eta})
	(\ell_\bot +i\frac{\ell_{''}}{\cosh\eta})} \ \ \ .
\end{equation}
This reveals the branch points for $\ell_{(\rm new \ frame)}$as a
function of the boosting rapidity $\eta$ given by 

\begin{eqnarray}
\eta&=&\makebox{arcosh}\left(\pm\frac{i\ell_{''}}{\ell_\bot}\right) =
i\frac{\pi}{2}\pm\makebox{arcsinh}\left(\ell_{''}/{\ell_\bot}\right) \ \ \ ,\\
\makebox{and }&&\nonumber\\
\eta&=&\makebox{arcosh}(0)=i\frac{\pi}{2}(\makebox{mod}\pi) \ \ \ .
\end{eqnarray}
The result of a boost by $\eta=i\pi$ may depend on which path of
$\eta$ from 0 to $i\pi$ one chooses.
Notice that we cannot let $\eta$ go along the pure imaginary axis
because it would hit the two branch points at $\eta=i\frac{\pi}{2}$ .
In fact each of the two factors 
$\sqrt{\ell_\bot\pm i\frac{\ell_{''}}{\cosh\eta}}$ 
has a branch point at $\eta=i\frac{\pi}{2}$.
Thus there is only a pole and no branch point for the $\ell_{(\rm new
\ frame)}$ expression.
Therefore it does not matter to the endresult of $\ell_{(\rm new \
frame)}$ on which side of the  singularity at $\eta=i\frac{\pi}{2}$
the passage has been performed to reach the endpoint $\eta=i\pi$.
So we can decide to make the path go by deforming to either side by
infinitesimal amounts $\pm\epsilon$ around this pole and the end
result will be the same.
We take the path outside the branch point singularities at 
$\eta=i\frac{\pi}{2}\pm \sinh^{-1}\frac{\ell_{''}}{\ell_\bot}$.
But these are at finite distance from the imaginary axis as long as
$\ell_{''}\neq 0$.
In this way we have seen that for $\ell_{''}\neq 0$ the most direct
path is the one connecting those two true branch points which is
almost the imaginary axis except for $\pm\epsilon$ deformation.
If we choose this path it is easy to show that a sign of
$\ell_{(\rm new \ frame)}$ gets changed in the sense that 

\begin{eqnarray}
\ell_{(\rm new \ frame)}(\eta=i\pi)&=&
	-\ell_{(\rm new \ frame)} (\eta=0)\nonumber\\
	&=&-\ell
\end{eqnarray}
Since all 4 vectors change sign by going to $\eta=i\pi$ the stick
should lie between $\stackrel{\rightharpoonup}{x}=0$ and
$\stackrel{\rightharpoonup}{x}=(-\ell_{''} \ , \ -\ell_\bot \ , 0)$ .
If we write pure imaginary $\eta$ as $\eta=i\hat\eta$ where $\hat\eta$ 
is real, we have $\cosh\eta=\cos\hat\eta$.
Thus the full radicant

\begin{equation}
\ell^2_\bot+\frac{\ell^2_{''}}{\cosh^2\eta}=
	\ell^2_\bot+\frac{\ell^2_{''}}{\cos^2\hat\eta}
\end{equation}
is positive and lies on the purely imaginary axis with
$\eta=i\hat\eta$.
If the radicant remains real and nonzero 
the square root $\eta_{(\rm new \ frame)}$ would stay positive say, by 
continuity.
Only when we make the detour with $\pm\epsilon$ is there a possibility 
to let $\eta_{(\rm new \ frame)}$ go into the complex quantity and
thus it may possibly change sign
when returning to the imaginary axis.

In fact since there is the pole in 
$\ell_{(\rm new frame)}$ at $\eta=i\frac{\pi}{2}$ the length
$\ell_{(\rm new \ frame)}$ changes sign when $\eta$ passes
from 
$\eta=i\left(\frac{\pi}{2}-\hat\epsilon\right)$ to
$\eta=i\left(\frac{\pi}{2}+\hat\epsilon\right)$ along a little detour
away from the imaginary axis.
Thus the result is that along the path described above there is indeed 
a sign change on the $\ell_{(\rm new \ frame)}$.

This means that for other pairs of particles we could also get such a
sign shift for the distance between the particles $r_{ik}=\ell_{(\rm new \ 
frame)}$.
We then could achieve both the operations A) and B) as an analytic
continuation of a Lorentz transformation.

We will not prove here the fact that one can always find a path along
which shifts sign on such distance.
However, we showed that by choosing as the suggestive shortest path of 
analytic continuation the one along the imaginary $\eta$-axis we get
the sign shift. 

So if it should at all be possible not to switch the signs of
distances by making an analytic continuation to $\eta=i\pi$, it would
at least be a more complicated and/or far away path.


\section{Relation to the usual CPT-theorem}

As we have described in the previous section we think that the new
CPT-like theorem holds in the naive vacuum world where there are a
finite number of particles.
However, if we are not afraid of infinities we may apply it on the
correct vacuum where all the fermion negative energy states are filled 
and for the bosons one particle is removed from each negative energy
single particle states.
In this case, however, the state is brought into another world in
which the positive energy states are all filled for fermions and for
bosons one particle is removed from each positive energy single
particle states.
Thus it is not a practical symmetry for the real world described by
the correct vacuum.
The CPT-like symmetry brings the real world across the figure with the 
four vacua as is illustrate in Fig.1, in much the same way as the
usual CPT-theorem brings the naive vacuum across the figure.

If we take a point of view that we apply both new and old CPT-theorems 
on all the four vacua and associated worlds, we may consider the
composite operation ;
\begin{eqnarray}
&&\makebox{(CPT-like \ symmetry)}\cdot\makebox{(usual \ CPT)}\nonumber\\
&=& 
\makebox{particle-hole \ exchange \ with \ analytic \ continuation  }
\nonumber
\end{eqnarray}
In fact the composite operation is the symmetry under the following
combined set of three operations :

\begin{enumerate}
\renewcommand{\labelenumi}{\arabic{enumi})}
\item  Particle hole exchange ; it replaces every particles by
corresponding holes which is equivalent to the $-1$ particle in the
same single particle state.
\item  Perform the same analytic continuation to all the particles so as
to bring all the relative distances $r_{ik}$ of them to change sign
\item  Complex conjugation of the wave function.
This operation is a symmetry in the sense that it brings the energy and 
momentum to the ones with opposite sign.
\end{enumerate}

It should be noted that this composite operation is antiunitary.
This symmetry can be derived under very general conditions that
consist mainly of the locality of interactions and some remnant of
Lorentz symmetry.
The point is that in order to prove the sign change of the
interactions by going to the other sheet in the distances
$r_{ik}\dot\psi$ in the boundary conditions eq.(\ref{10.1}) should be shown to
behave as an odd power of ${r_{ik}}^{-1}$.

Certainly one can claim that the particle hole exchange theorem may be 
proven by combining our arguments in section 10 for the proof of the
CPT-like theorem in the naive vacuum with the usual CPT theorem.
However we believe that the proof could be made under much weaker
assumptions.


\section{Conclusions}

We have put forward an attempt to extend also to bosons the idea of
Dirac sea for fermions.
We first consider one second quantization called the naive vacuum
world in which there exist a few positive and negative energy fermions 
but no Dirac sea yet.
This first picture of the naive vacuum world model is very bad with
respect to physical properties in as far as no bottom in the
Hamiltonian.
For bosons this naive vacuum is even worse physically because in
addition to negative energies without bottom a state with an odd number 
of negative energy bosons has negative norm square.
There is no real Hilbert space but only an indefinite one.
At this first step of the bosons the inner product for the Fock space
is not positive definite.
Thus this first step is completely out from the phenomenological point 
of view for the bosons as well as for the fermions.
For the bosons for two major reasons : negative energy and negative
norm square.

However, from the point of view of theoretical study this naive vacuum 
world at the first step is very attractive because the treatment of a
few particles is quantum mechanics rather than quantum field theory.
Furthermore by locality the system of several particles becomes free
in the neighborhood of almost all configurations except for the case
that some particles meet and interact.
We encourage the use of this theoretically attractive first stage as a 
theoretical playground to gain understanding of the real world, the
second stage.

In the present article we studied the naive vacuum world at first
stage.
We would like to stress the two major results in the following :

\begin{enumerate}
\renewcommand{\labelenumi}{\arabic{enumi})}
\item We found a CPT-like symmetry.
A reduced form of strong reflection provides an extra transformation
that is an analytic continuation of the wave function onto another
sheet among $2^{\frac{1}{2}N(N+1)}$ ones for the wave function of 
the $N$ particle system.
The sheet structure occurs because $r_{ik}$ is a square root so that
it has 2 sheets.
For each of the $\frac{1}{2}N(N+1)$ pairs of particles there is a
dichotomic choice of sheet so that it gives $2^{\frac{1}{2}N(N+1)}$
sheets.
\item We found the main feature of the wave functionals for the bosons. 
They are derivatives of $\delta$-functionals of the complex field
multiplied by polynomials in the complex conjugate of the field.
These singular wave functionals form a closed class when acted upon by
polynomials in creation and annihilation operators.
Especially we worked through the case of one pair of a single particle 
state with a certain momentum and the one with the opposite momentum.
\end{enumerate}

The main point of our present work was to formulate the transition
from the naive vacuum of the first stage into the next stage of the
correct vacuum.
For fermions it is known to be done by filling of the negative energy
states which is nothing but filling up the Dirac sea.
The corresponding procedure to bosons turned out to be that from each
negative energy single particle states one boson is removed, that is
minus one boson is added.
This removal cannot be done quite as physically as the adding of a
fermion, because there is a barrier to be crossed.

We did this by studying the harmonic oscillator corresponding to a
single particle boson state.
We replaced the usual Hilbert norm requirement of finiteness by the
requirement of analyticity of the wave function in the whole complex
$x$-plane except of $x=\pm\infty$.
The spectrum of this extended harmonic oscillator or the harmonic
oscillator with analytic wave function has an additional series of
levels with negative energy in addition to the usual one.
The wave functions with negative energy are of the form with Hermite
polynomials times $e^{\frac{1}{2}(\beta x)^2}$.

We note that there is a barrier between the usual states and the one
with negative excitation number because annihilation or creation
operators cannot cross the gap between these two sectors of states.
The removal of one particle from an empty negative energy state
implies crossing the barrier. 
Although it cannot be done by a finite number of interactions
expressed as a polynomial in creation and annihilation operators we
may still think of doing that.
Precisely because of the barrier it is allowed to imagine the
possibilities that negative particle numbers could exist without
contradicting with experiment.

Once the barrier has been passed to the negative single particle
states in a formal way the model is locked in and those particles
cannot return to the positive states. 
Therefore it is not serious that the correct vacuum for bosons get a
{\it higher} energy than the states with a positive or zero number of
particles in the negative energy ones.


\section{Acknowledgment}

We would like to thank J. Greensite for useful discussions 
on string field theories.
Main part of this research was performed at YITP while one of us
(H.B.N.) stayed there as a visiting professor.
H.B.N. is grateful to YITP for hospitality during his stay.
M.N. acknowledges N.B.I for hospitality extended to him during his
visit.


\end{document}